

\input harvmac


\def\NSNS{{$NS\otimes NS $}}
\def\RR{{$R\otimes R$}}

\def\II{\relax{{\rm I}\kern-.10em {\rm I}}}
\def\twoa{{\II}{\rm A}}

\def\ele{\relax{1\kern-.10em 1}}

\def\g(#1){\Gamma(#1)}

\def\C|#1{{\cal #1}}
\def\(#1#2){(\zeta_#1\cdot\zeta_#2)}

\def\Tr{{\rm Tr}}

\def\half{  {1 \over 2}}
\def\quart{ {1 \over 4}}

\def\hmu{{\hat \mu}}
\def\hnu{{\hat \nu}}

\def\hrho{{\hat \rho}}
\def\hsigma{{\hat \sigma}}
\def\hchi{{\hat \chi}}
\def\hm{\hat{m}}
\def\hn{\hat{n}}

\def\calS{{\cal S}}
\def\calR{{\cal R}}
\def\calV{{\cal V}}


\def\p{\partial}

\def\g{\gamma}

\def\tde{\tilde\delta}
\def\pt{\tilde\psi}
\def\dtau{{d\over d\tau}}
\def\ast{C^{(3)}}

\def\expo{e^{-ik\cdot X}}
\def\sqrtp{\sqrt{p^+}}
\def\thbar{\bar{\theta}}
\def\eqs{\eqalign}

\tolerance=10000
\def\xxx#1 {{hep-th/#1}}
\def\lr { \lref}
\def\npb#1(#2)#3 { Nucl. Phys. {\bf B#1} (#2) #3 }
\def\rep#1(#2)#3 { Phys. Rept.{\bf #1} (#2) #3 }
\def\plb#1(#2)#3{Phys. Lett. {\bf #1B} (#2) #3}
\def\prl#1(#2)#3{Phys. Rev. Lett.{\bf #1} (#2) #3}
\def\physrev#1(#2)#3{Phys. Rev. {\bf D#1} (#2) #3}
\def\ap#1(#2)#3{Ann. Phys. {\bf #1} (#2) #3}
\def\rmp#1(#2)#3{Rev. Mod. Phys. {\bf #1} (#2) #3}
\def\cmp#1(#2)#3{Comm. Math. Phys. {\bf #1} (#2) #3}
\def\mpl#1(#2)#3{Mod. Phys. Lett. {\bf #1} (#2) #3}
\def\ijmp#1(#2)#3{Int. J. Mod. Phys. {\bf A#1} (#2) #3}
\def\jhep#1(#2)#3{JHEP {\bf #1} (#2) #3}
\def\adv#1(#2)#3{ Adv. Theor. Math. Phys. {\bf #1} (#2) #3}
\def\hep#1{{\tt hep-th/#1}}


\lr\goroff{M. Goroff and J.H. Schwarz, {\it D-dimensional gravity in
the light-cone gauge}, \plb127(83)61.}
\lr\sugra{E. Cremmer, B. Julia and  J.  Scherk, {\it Supergravity
theory in eleven dimensions},  \plb76(1978)409. }
 \lr\worldline{Z.~Bern and D.A.~Kosower,
{\it A new approach to one loop calculations in gauge theories},
Phys. Rev. {\bf D38} (1988) 1888 ; M.J.~Strassler,
{\it Field theory without Feynman diagrams: One loop effective actions},
Nucl. Phys. {\bf B385} (1992) 145 ,
hep-ph/9205205. }
\lr\wittena{E.~Witten,
{\it String theory dynamics in various dimensions},
Nucl. Phys. {\bf B443} (1995) 85 ,
hep-th/9503124.}
\lr\redMIIA{I.C.~Campbell and P.C.~West,
{\it N=2 D = 10 Nonchiral Supergravity And Its Spontaneous
Compactification},
Nucl. Phys. {\bf B243} (1984) 112 ; M.~Huq and M.A.~Namazie,
{\it Kaluza-Klein Supergravity In Ten-Dimensions},
Class. Quant. Grav. {\bf 2} (1985) 293 . }
\lr\gsc{M.B.~Green and J.H.~Schwarz,
{\it Covariant Description Of Superstrings},
Phys. Lett. {\bf 136B} (1984) 367 ; { \it Properties Of The Covariant
Formulation Of Superstring Theories},
Nucl. Phys. {\bf B243} (1984) 285.}
\lr\greenseib{M.B.~Green and N.~Seiberg,
{\it Contact Interactions In Superstring Theory},
Nucl. Phys. {\bf B299} (1988) 559 .}
\lr\berghull{hep-th/9504081 Bergshoeff, Hull and Ortin}
\lr\ggv{M.B.  Green, M.  Gutperle and P.  Vanhove, {\it One loop in eleven dimensions},  \plb409(1997)177, hep-th/9706175.}
\lr\ggk{M.B.  Green, M. Gutperle and H.  Kwon, {\it Sixteen fermion and related terms in M theory on $T^2$}, Phys. Lett. {\bf B421} (1998)  149, hep-th/9710151.}
\lr\gsl{D.J.~Gross and J.H.~Sloan,
{\it The Quartic Effective Action For The Heterotic String},
Nucl. Phys. {\bf B291} (1987) 41 .}
\lr\kallosha{R.~Kallosh,
{\it Covariant quantization of D-branes},
Phys. Rev. {\bf D56} (1997) 3515,
hep-th/9705056.}
\lr\howewest{P.  Howe and P. West, {\it The Complete N=2, D = 10 Supergravity }, Nucl. Phys. {\bf B238} (1984) 181.}
\lr\sav{M.B.~Green and S.~Sethi,
{\it Supersymmetry constraints on type IIB supergravity},
Phys. Rev. {\bf D59} (1999) 046006,
hep-th/9808061.}
\lr\kehagias{A.~Kehagias and H.~Partouche,
{\it The Exact quartic effective action for the type IIB superstring},
Phys. Lett. {\bf B422} (1998)  109,
hep-th/9710023.}
\lr\mgmbgd{M.B.~Green and M.~Gutperle,
{\it Effects of D instantons},
Nucl. Phys. {\bf B498} (1997)  195,
hep-th/9701093.}
\lr\bern{
Z.~Bern, L.~Dixon, D.C.~Dunbar and D.A.~Kosower,
{\it Advanced techniques for multiparton loop calculations: A Minireview},
hep-ph/9706447 and references therein. }
\lr\deser{S. Deser et al  hep-th/9812136, hep-th/9805205  }
\lr\cremmer{E. Cremmer and  S. Ferrara, {\it Formulation Of Eleven-Dimensional Supergravity In Superspace}, \plb91(1980)61.}
\lr\schwarzaspinwall{J.H. Schwarz, {\it The Power of M Theory},
Phys. Lett. {\bf B367} (1996) 97; P.S.~Aspinwall,
{\it Some relationships between dualities in string theory},
Nucl. Phys. Proc. Suppl. {\bf 46} (1996) 30,
hep-th/9508154.}
\lr\mbgconf{M.B.~Green,
{\it Connections between M theory and superstrings},
Nucl. Phys. Proc. Suppl. {\bf 68} (1998) 242,
hep-th/9712195.}
\lr\tsru{J.G.~Russo and A.A.~Tseytlin,
{\it One loop four graviton amplitude in eleven-dimensional supergravity},
Nucl. Phys. {\bf B508} (1997) 245,
hep-th/9707134.}
\lr\berntwo{Z.~Bern {\it et al.},
{\it On the relationship between Yang-Mills theory and gravity and its
                  implication for ultraviolet divergences},
Nucl. Phys. {\bf B530} (1998) 401,
hep-th/9802162;  Z.~Bern, L.~Dixon, M.~Perelstein and J.S.~Rozowsky,
{\it Multileg one loop gravity amplitudes from gauge theory},
hep-th/9811140.}
\lr\banks{T.~Banks, W.~Fischler, S.H.~Shenker and L.~Susskind,
{\it M theory as a matrix model: A Conjecture},
Phys. Rev. {\bf D55} (1997) 5112,
hep-th/9610043.}
\lr\gv{M.B.~Green and P.~Vanhove,
{\it D instantons, strings and M theory},
Phys. Lett. {\bf B408} (1997) 122,
hep-th/9704145.}
\lr\plefka{J. Plefka, M. Serone and A. Waldron, {\it D=11 SUGRA as
the Low Energy Effective Action of Matrix Theory: Three Form
Scattering}, \jhep9811(1998)010, \hep9809070. }
\lr\piolinea{B.~Pioline and E.~Kiritsis,
{\it U duality and D-brane combinatorics},
Phys. Lett. {\bf B418} (1998) 61,
hep-th/9710078; {\it On $R^4$ threshold corrections in IIb string
theory and (p, q) string instantons},
Nucl. Phys. {\bf B508} (1997)  509,
hep-th/9707018.}
\lr\berkovits{N.~Berkovits and C.~Vafa,
{ \it Type IIB $R^4 H^{4g-4}$ conjectures},
Nucl. Phys. {\bf B533} (1998) 181,
hep-th/9803145.}
\lr\dewit{ B. de Wit, K. Peeters and J. Plefka, {\it Superspace
geometry for supermembrane backgrounds}, \npb532(1998)99, \hep9803209.}
\lr\gs{ M. B. Green and J. H. Schwarz, {\it Supersymmetric dual string
theory (III); loops and renormalization}, \npb198(1982)441.}
\lr\green{ M. B. Green, {\it Interconnections between type II
superstrings, M theory and N=4 supersymmetric Yang--Mills},
\hep9903124.}
\lr\juan{J.  Maldacena, {\it   The large $N$ limit of
superconformal field
theories and supergravity}, \adv2(1998)231, \hep9711200.}
\lr\gkpw{S.S. Gubser,  I.R.  Klebanov and A.M.  Polyakov, {\it
Gauge theory
correlators from non-critical string theory}, \plb428(1998)105,
\hep9802109 \semi 
E.  Witten, {\it Anti de Sitter Space and Holography}, \adv2(1998)253,
 \hep9802150. }


\noblackbox
\baselineskip 14pt plus 2pt minus 2pt
\Title{\vbox{\baselineskip12pt
\hbox{hep-th/9907155}
\hbox{DAMTP-1999-92}
\hbox{PUPT-1880}}}
{\vbox{\centerline{Light-cone Quantum Mechanics}\bigskip
\centerline{of the Eleven-dimensional
 Superparticle }}}
 \centerline{ Michael B. Green$^{1}$,  Michael Gutperle$^{2}$ and
Hwang-h. Kwon$^{1}$ }
\medskip
\centerline{$^{1}$DAMTP, Silver Street, Cambridge CB3 9EW, UK}
\centerline{$^{2}$Physics Department, Princeton University, Princeton
NJ 08544, USA}
\medskip
\centerline{m.b.green@damtp.cam.ac.uk,
gutperle@feynman.princeton.edu, h.kwon@damtp.cam.ac.uk}
 \bigskip
 \medskip

\centerline{\bf Abstract}

The linearized interactions of  eleven-dimensional supergravity are
obtained in a manifestly supersymmetric light-cone gauge formalism. 
These vertices are  used to  calculate certain  one-loop processes
involving external gravitini, antisymmetric three-form potentials and gravitons,
 thereby determining some protected terms in the effective action of
M-theory  compactified on a two-torus.

\noblackbox
\baselineskip 14pt plus 2pt minus 2pt

\Date{ July 1999}

\hoffset=-.3 in

\lineskip= 0.5pt


\newsec{ Introduction }
Classical eleven-dimensional supergravity \sugra\  is the long
wavelength or low energy  limit of M-theory \wittena.    In
a number of  papers \refs{\ggv,\ggk,\mbgconf,\tsru}
it has been  shown that certain one-loop
quantum calculations in compactified
eleven-dimensional supergravity generate terms in the effective
M-theory action that arise  in string theory  as perturbative and
non-perturbative effects.  These loop  calculations would have been
very  complicated using  standard   Feynman rules for the component
fields,  in which the many cancellations between different
contributions are not at all apparent.  Such  cancellations would be
natural in a covariant eleven-dimensional   superspace formalism, but such a
formalism only exists for the on-shell theory \refs{\cremmer}.
In  the absence of
 useful eleven-dimensional covariant superspace Feynman rules, these
one-loop calculations made use of a supersymmetric  light-cone
gauge. in which  sufficient supersymmetry is
manifest to streamline the calculations.  The purpose of this paper is
to obtain the light-cone gauge Feynman rules that were used in the earlier
papers and illustrate their use by evaluating some further one-loop amplitudes.

We will start from the quantum mechanical description of the massless
eleven-dimensional superparticle to obtain vertex operators that
describe interacting particles in linearized approximation. This will
be sufficient to evaluate the one-loop
Feynman diagrams  by integrating over
 the  world-lines of the circulating particles, with the vertex operator
insertions representing the interactions with the external particles.
This approach is modeled
on the standard methods for evaluating string theory diagrams.   It has also
 proved to be an efficient method \refs{\worldline} for calculating
radiative corrections to Yang--Mills theories of relevance to the
Standard Model \refs{\bern} and S-matrix elements in $N=8,d=4$ supergravity
\berntwo.  In the case of the four-graviton amplitude the loop 
calculation reduces  to the strikingly simple form of a 
simple kinematic factor multiplying a scalar 
field theory box diagram.

In section 2 the  global and local symmetries of the
eleven-dimensional superparticle action   will be elucidated and the
light-cone  superspace
quantum mechanics described.  The light-cone superspace form of the
vertex operators that describe the
cubic interactions  of the component fields
will be introduced  in section
 3.     As with the analogous superstring vertices these describe the
emission of on-shell particles.  In this case these comprise the
graviton, gravitino and three-form potential with vertices that
will be denoted $V_h$, $V_\psi$ and $V_{C^{(3)}}$,  respectively.
The form of these vertex operators is  uniquely determined  by
the requirement that they  transform appropriately under
the 32-component supersymmetry transformations, which form an
$SO(10,1)$ spinor.  In the
light-cone gauge these supersymmetries divide into sixteen
linearly realized transformations and sixteen that are realized
nonlinearly.   The linear supersymmetry transformations are
sufficiently simple to be checked completely.  In order to determine
the  complete expressions for the vertices
 the only nonlinear supersymmetry
transformation that needs to be checked is that  of the graviton
vertex. The nonlinear transformations of the other vertices are
complicated and will not be considered. Certain total
derivatives with respect to the world-line time parameter arise
in the closure for the supersymmetry algebra which determine the
time variations of the cubic interaction contributions to the
light-cone supercharges. Closure of the supersymmetry algebra
also generates higher-order interaction terms involving arbitrary
powers of the superfields, but these will not be considered here.
 In section 4  dimensional reduction of the linearized
eleven-dimensional theory
on a circle  is  shown to reproduce  the point particle
limit of the corresponding linearized IIA superstring
theory.

 Section 5 will describe the use of these vertex operators to
calculate  certain classes of 
one-loop amplitudes in eleven-dimensional supergravity compactified on
a circle or a two-torus.  The
particular processes that we will consider are ones that correspond to 
interactions in the effective action that are integrals 
over half the on-shell superspace and are very strongly constrained by
supersymmetry.  We will argue that 
the one-loop amplitudes for such  `protected' processes may be
calculated completely by using the linearized (cubic) interaction
vertices and do not  recieve contributions from higher order contact
terms.   
 In part this will fill in details used in
obtaining the  one-loop results
described in  \refs{\ggv,\ggk} where 
connections to  perturbative and nonperturbative D-instanton induced
terms in type  IIB string theory  \refs{\mgmbgd} were made.
 In addition some further amplitudes will be evaluated that lead to
terms in the IIB effective action involving the 
third-rank and self-dual
fifth-rank field strengths. 


\newsec{The massless eleven-dimensional superparticle in light-cone gauge}
The configuration space of an eleven-dimensional superparticle has
eleven bosonic coordinates, $X^\hmu$ (${\hmu}=0,1,\cdots,9,11$), which
form a  $SO(10,1)$ vector and 32 fermionic Grassmann coordinates,
$\Theta^{\hat A}$ ($\hat A = 1, \dots, 32$), that form a Majorana
$SO(10,1)$ spinor.   The action for such a particle is given by
\eqn\what{S_{particle}= {1\over 2} \int d\tau {1\over e}\Pi^{\hmu}\Pi_{\hmu}}
where
\eqn\pidef
{\Pi^{\hmu}=\dot{X}^{\hmu}-i\overline{\Theta}\Gamma^{\hmu}\dot{\Theta},}
 where the matrices $\Gamma^{\hmu}$ are $32 \times 32$-component Dirac
matrices.
The equations of motion that follow from this action are
\eqn\motdef{
  \Pi^2= 0 \qquad
\dot{\Pi}^{\hmu}= 0 \qquad
\Gamma^{\hmu}\Pi_{\hmu} \dot{\Theta}= 0\, .}
The action is invariant under global Poincar\'e and supersymmetry
transformations,
\eqn\globdeef{
\delta_\alpha\Theta=\alpha,\quad \delta_\alpha X^{\hmu}=
i\bar{\alpha}\Gamma^{\hmu}\Theta, }
where $\alpha^{\hat A}$ is a constant majorana spinor parameter,
as well as local reparameterizations and kappa symmetry transformations,
\eqn\kapdef{
  \delta_\kappa \Theta = i \Gamma^{\hmu}\Pi_{\hmu}\kappa, \quad \delta_\kappa
  X^{\hmu}=i\overline{\Theta} \Gamma^{\hmu} \delta_\kappa\Theta, \quad
 \delta_\kappa e = 4e\,\dot{\overline{\Theta}}\kappa \; , }
where the parameter $\kappa^{\hat A}(\tau)$ is a Majorana spinor and a
world-line scalar density.

There is no obvious way of formulating  the quantum mechanics of this
system covariantly   but it is straightforward  to quantize
the system in the light-cone gauge.   This is defined by using the
reparameterization invariance and kappa symmetry to choose
\eqn\lcdeef{X^+= x^+ + p^+\tau,\qquad \Gamma^+\Theta =0 ,}
where the light-cone coordinates are defined by\foot{With this choice
of light-cone directions the conventional coordinates of the IIA
theory will be obtained by compactifying the $X^{11}$ direction.}
\eqn\lccon{
X^+= {1\over \sqrt{2}}(X^0+X^9),\qquad   X^-= {1\over
\sqrt{2}}(X^0-X^9) }
and
\eqn\lcgam{
\Gamma^+ ={1\over \sqrt{2}}(\Gamma^0+\Gamma^9),\qquad   \Gamma^-= {1\over
\sqrt{2}}(\Gamma^0-\Gamma^9).
}
The matrix $\Gamma^+$ projects onto a $16 \times 16$ subspace spanned
by $SO(9)$ Majorana spinors.
In this gauge the action is expressed in terms  of the transverse
coordinates, $X^I(\tau)$ ($I=1,\cdots,8,11$), and the 16-component
$SO(9)$ spinor, $\calS^A$ ($A=1, \dots, 16$),  so that
\eqn\action{
{\cal S}_{l.c.}= \int d\tau\big(\half (\dot X^I)^2 -
i\calS\dot\calS\;\big) .}
  The equations of motion, $\partial\dot X^I /\partial \tau =0
=\partial\calS^A/\partial \tau  $  imply that both the momentum
operator, $p^I = \dot X^I$ and the fermionic operator  $\calS^A$ are
constant.  The (anti)commutation relations that follow from
 the Poisson brackets are
\eqn\comrel{[X^I,  p^J]= i\delta^{IJ}, \qquad \{\calS^A,\calS^B\}=
\delta^{AB} .}

 The   32 components of the $SO(10,1)$ space-time supersymmetry
 decompose into   a ${\bf 16}_+$ of $SO(9)\times U(1)$ that is
 linearly realized and a ${\bf 16}_-$  that is nonlinearly
 realized.
The linearly realized supersymmetries  are those components that
 satisfy  $\Gamma^+\alpha=0$   and will be associated with a spinor
 parameter $\eta$.  They are generated by the $S^A$ and
their action  on the coordinates is given by
\eqn\lin{
\delta X^I = 0, \qquad \delta\calS = \sqrt{p^+}\eta.
}
The nonlinearly realized supersymmetries  will be associated with a
second 16-component spinor parameter $\epsilon$  associated with the
components that satisfy
 $\Gamma^-\alpha=0$ and act  as
\eqn\nlin{
\tilde\delta X^I = -{2\over \sqrt{p^+}}\epsilon\gamma^I\calS, \qquad
\tilde\delta\calS ={i\over \sqrt{p^+}} \dot X^I\gamma^I\epsilon
}
where $\gamma^I$  is a $16 \times 16$ $SO(9)$ gamma matrix and the
$\tilde{}$ will  be used to distinguish the nonlinearly realized
symmetries from  the linearly realized ones.

The physical states consist of 44 transverse symmetric  traceless tensor
states $|IJ\rangle$ (with $|IJ\rangle = |JI\rangle$ and
$|II\rangle =0$)
representing the graviton, 128 gamma-tracelss
spinor-vector states $|AI\rangle$ (with $\Gamma^I_{AB}|AI\rangle =0$)
 representing the gravitino, and 84
antisymmetric tensor states $|LMN\rangle$ representing the three-form
potential. These states form a representation of the linear
supersymmetries
generated by the action of the sixteen components of ${\cal S}^A$,
\eqn\statetr{\eqs{
{\cal S}^A |IJ\rangle =&
\,\Gamma^I_{AB}|BJ\rangle+\Gamma^J_{AB}|BI\rangle\cr
{\cal S}^A |BI\rangle =&\, \quart\Gamma^J_{AB}|IJ\rangle
+{1\over 72}\big(\Gamma^{ILMN}+6\delta^{IL}\Gamma^{MN}\big)|LMN\rangle \cr
{\cal S}^A |LMN\rangle =& \, \Gamma^{LM}_{AB}|BN\rangle
+\Gamma^{MN}_{AB}|BL\rangle +\Gamma^{NL}_{AB}|BM\rangle.
}}


\subsec{Supersymmetry in the light-cone gauge} The Lagrangian of
eleven-dimensional supergravity \sugra\ contains three fields: the
graviton $h_{\hmu\hnu}$, the gravitino $\Psi^{\hat A}_{\hmu}$ and
the three-form potential $C_{\hmu\hnu\hrho}$.
 The covariant equations of motion  imply   $k^\hmu k_\hmu =0$
together with the  physical state conditions,
\eqn\condef{
k_{\hmu} h^{\hmu}_{\ \hnu} =\half k_{\hnu} h^{\hmu}_{\ \hmu}, \qquad
\Gamma^\hmu \Psi_\hmu =   k_{\hmu} \Psi^{\hmu}
=  \Gamma \cdot k \Psi_\hmu = 0 ,\qquad    k^{\hmu}
\ast_{\hmu\hnu\hrho} =0 \, .}

 The  light-cone gauge  is reached by using the reparameterization
 invariance, local supersymmetry and the local symmetry associated
 with  the three-form potential to impose the conditions
\eqn\lcfields{h_\hmu^{\ +} =0, \qquad \Psi^{\hat A +} =0, \qquad
C_{\hmu\hnu}^{\ \ +} =0.}
 As usual, the light-cone vertex operators will  have a particularly
 simple form in  a  frame in which $k^+ =0$, which is attainable in
 general  when there are few enough external particles. It is also
 convenient to take  $k^-$ to be finite so that,  with the  condition
 $k^\hmu k_{\hmu} =0$,
the transverse  momenta satisfy $k^I k_I =0$, which is only possible
for complexified momenta.
  The physical momenta can then be reached
by analytic continuation.  This kinematic set-up  has proved useful in
 superstring calculations and will be general enough for our purposes.
 In this case the conditions \condef\  become
\eqn\conlc{
k_I h^I_{\  \hmu} =\half k_{\hmu}h^I_{\ I}, \quad k_I \Psi^I =0=
(\Gamma^J  k_J - \Gamma^+ k^-) \Psi_\hmu ,\quad \Gamma^I\Psi_I=
\Gamma^+\Psi^-, \quad k^{I} \ast_{I\hmu\hnu} = 0 .}

 The ${}^-$ components are determined
by the constraints \condef\  for non-zero $k^+$ but are
unrestricted  when $k^+=0$. In writing
\conlc\ we have assumed that  the components,
\eqn\indef{h_\hmu^{\ -}, \qquad  \Psi^{-A},  \qquad C_{IJ}^{\ \ -},}
are non-infinite.    Another condition that follows from \condef\ when
$k^+ \ne 0$ is the tracelessness condition, $h_I^{\ I}
=0$, from which it follows that  the physical state condition for the graviton
in \conlc\ is,
\eqn\gravcon{  k_I h^I_{\  \hmu} =0.
}
When $k^+=0$,  the light-cone tracelessness condition
does not  follow
from \condef\ but it can be imposed by hand.

 The physical  fields are classified in  representations
of $SO(9)$.   Thus $h_{IJ}$ is a  traceless symmetric second-rank
tensor,  while $h_I^{\ -}$ is a vector and $h^{--}$ is a scalar.  The
components of the three-form potential are  $C_{IJK}$  and
$C^{\ \  -}_{IJ}$. The  gravitino decomposes into two $SO(9)$ parts,
\eqn\psidefs{\Psi^{\hat A}_{\hmu} \equiv P^+ \Psi^{\hat A}_{\hmu} + P^-
\Psi^{\hat A}_{\hmu} = (\psi^A_{\hmu}, \tilde \psi^A_{\hmu}),}
where $P^\pm = \half \Gamma^\pm \Gamma^\mp$. The transverse
components $\psi^A_I, \pt^A_I$ are two spinor-vectors  while
$\psi^{-A}, \pt^{-A}$ are two spinors.
These satisfy the constraints,
\eqn\spincons{\eqalign{\gamma^I_{AB}\tilde \psi_I^B = 0,  \quad &
\gamma^I_{AB} \psi_I^B = \tilde \psi^{-A},  \quad k^I \psi_I^A = 0,
\quad k^I \tilde \psi_I^A=0,
\cr  \gamma^I k^I \tilde \psi_\hmu &=0, \quad \gamma^I k^I \psi_\hmu
 = k^- \tilde \psi_\hmu .
}}


 The covariant
supersymmetry transformations of the component fields are  given
by\foot{ Numerical
factors differ slightly from \refs{\sugra}. A similar normalization has been
taken in \dewit.}
\eqn\cosusy{
\delta h_{\hmu\hnu} = \bar{\alpha}\Gamma_{(\hmu}\Psi_{\hnu)}, \quad
\delta \Psi_{\hmu} = D_{\hmu}(\hat{\omega})\alpha +
T_{\hmu}^{\ \hnu \hrho \hsigma \hchi} \alpha
\hat{F}_{\hnu\hrho\hsigma\hchi}, \quad
\delta C^{(3)}_{\hmu\hnu\hrho} = {3\over 2} \bar{\alpha}
\Gamma_{[\hmu\hnu} \Psi_{\hrho]} ,}
where the linearized covariant derivative is defined by
\eqn\linearized{
  D_{\hmu}(\hat{\omega})\alpha = \big(k_{\hmu}
+k_{[\hrho}h_{\hsigma]\hmu}\Gamma^{\hrho\hsigma} \big)\alpha
}
and the super-covariant $\hat F_{\hmu\hnu\hrho\hsigma} $ is simply the
field strength $ F_{\hmu\hnu\hrho\hsigma} = 4 k_{[\hmu}
C^{(3)}_{\hnu\hrho\hsigma]}$. The tensor $T$ is defined by
\eqn\tdef{
  T^{\hmu\hnu\hrho\hsigma\hchi}={1\over 72}\left(
  \Gamma^{\hmu\hnu\hrho\hsigma\hchi}-8
\Gamma^{\hrho\hsigma\hchi}\eta^{\hmu\hnu}\right) ,}
where   $\eta^{\hmu\hnu}$ is the Minkowski metric.

In the light-cone gauge  the  supersymmetry transformations  \cosusy\
must be accompanied by compensating gauge transformations in order to
preserve the gauge conditions  \lcfields.     The compensating
reparameterizations are determined by
\eqn\comgrav{
\delta h_I^{\ +} = \half\bar{\alpha}(\Gamma^+ \Psi^I + \Gamma^I \Psi^+)
+k^+\xi^I  +k^I \xi^+  .}
so that the  light-cone gauge condition, $h_I^+ = 0$ is preserved  by choosing
\eqn\comppar{
\xi^I = -{1\over 2k^+}\bar{\alpha}\Gamma^+\Psi^I,    \qquad \xi^+ =0 \; . }
Hence the transverse components of the graviton transform in the
following way,
\eqn\lcgrav{\eqalign{
\delta h_{IJ} &= \bar{\alpha}\Gamma_{(I}  \Psi_{J)}  -{1\over
k^+}\bar{\alpha}\Gamma^+k_{(I} \Psi_{J)}  \cr
&=  \eta\g_{(I}\pt_{J)}
+\epsilon\g_{(I}\psi_{J)}-\sqrt{2}\epsilon{k_{(I}\pt_{J)}\over k^+} ,
}}
where $\psi_I, \pt_I, \eta$ and $\epsilon$ are $SO(9)$ Majorana spinors
defined by
\eqn\proj{\eqalign{
\psi_I &= \half \Gamma^+\Gamma^-\Psi_I , \qquad  \pt_I = \half
\Gamma^-\Gamma^+\Psi_I \cr
\eta &= \half \Gamma^+\Gamma^-\alpha,  \qquad \epsilon = \half
\Gamma^-\Gamma^+\alpha .
}}

The reparametrization transformations on $\Psi_I$ and $C_{LMN}^{(3)}$
vanish in the free-field limit and supersymmetry transformations
for these fields in \cosusy\ are to be compensated by gauge
transformations,
\eqn\compsi{
\delta\Psi_\hmu  = k_\hmu \beta,  \qquad \delta C^{(3)}_{\hmu\hnu\hrho} =
3k_{[\hmu}\upsilon_{\hnu\hrho]} ,}
where the gauge parameters $\beta$ and $\upsilon^{\hmu\hnu}$ are
an $SO(10,1)$ Majorana spinor and a two-form, respectively. Their compensating
gauge transformations are
\eqn\comptrs{
\beta= -\alpha -{1\over k^+}T^{+\hmu\hnu\hrho\hsigma}\alpha
F_{\hmu\hnu\hrho\hsigma}, \qquad
\upsilon_{IJ} = -{1\over k^+}\bar{\alpha}\Gamma^+_{\ [I} \Psi_{J]} ,
\qquad  \upsilon_I^+  =0  .}
Hence the transverse components of the gravitino transform in the
following way
\eqn\vargau{
\delta\Psi_I = D_I\alpha +{1\over
72} (\Gamma_I^{\  \hmu\hnu\hrho\hsigma} F_{\hmu\hnu\hrho\hsigma} -
8\Gamma^{\hmu\hnu\hrho} F_{I\hmu\hnu\hrho} )\alpha-k_I\alpha -
{k_I\over k^+} T^{+\hmu\hnu\hrho\hsigma}\alpha F_{\hmu\hnu\hrho\hsigma} ,}
and
\eqn\moreg{
\delta C^{(3)}_{LMN} = {3\over 2}\bar{\alpha}\Gamma_{[LM}\Psi_{N]}
-{6\over k^+}\bar{\alpha}k_{[L}\Gamma^+_M\Psi_{N]} \; . }
In terms of $SO(9)$ spinors,
\eqn\lcrs{\eqalign{
\delta\psi_I =& k_{[L}h_{M]I}\gamma^{LM}\eta +{1\over 72}(\gamma_I^{\
JLMN}F_{JLMN}+24\gamma^{LMN}k_L\ast_{IMN}-4\gamma^{LMN}k_I\ast_{LMN})\eta
\cr
&-k_{[+}h_{J]I}\gamma^J\epsilon   -{\sqrt{2}\over
72}\,{k_I\over k^+}\gamma^{JLMN}F_{JLMN}\;\epsilon +\cdots , \cr
\delta\pt_I =& -{\sqrt{2}\over 2}k^+h_{IJ}\gamma^J\eta
 +{\sqrt{2}\over
 18}k^+(\g^{ILMN}-6\g^{MN}\delta^{IL})\ast_{LMN}\;\eta\cr
&+ k_{[L}h_{M]I }\gamma^{LM}\epsilon
+{1\over 72}
(\gamma_I^{\
 JLMN}F_{JLMN}+24\gamma^{LMN}k_L\ast_{IMN}+4\g^{LMN}k_I\ast_{LMN})\;\epsilon
  \cr &+\cdots
}}
where $\cdots$ indicate  terms with longitudinal polarizations  and
\eqn\delcc{
\delta C^{(3)}_{LMN} = {3\over 2}\eta\gamma_{[LM}\pt_{N]}+ {3\over
2}\epsilon\gamma_{[LM}\psi_{N]}
-{6\sqrt{2}\over k^+}\epsilon k_{[L}\gamma_{M}\pt_{N]}  .}


\newsec{Supersymmetry and the vertex operators}

Before showing in some detail how the vertex operators are
determined by the requirement that they form a representation of
the supersymmetry algebra we will summarize the results. The
vertex operator describing the emission of an on-shell physical
field, $\Phi$, will be written in the form
\eqn\formvert{V_\Phi = U_\Phi e^{-ik \cdot X} = U_\Phi e^{-i k_I
X^I(\tau) } e^{ik^-p^+ \tau},} where the   prefactor $U_\Phi$
depends on the species of field.

The following notation will also be introduced,
\eqn\rdefine{\calR^{LMN} = {1\over 12}\calS\gamma^{LMN}\calS,\quad
\calR^{IL} = {1 \over 4}\calS\gamma^{IL}\calS\; . } The operators
$\calR^{IL}$  form a representation of the transverse $SO(9)$
algebra and are  transverse angular momentum operators. Extensive
use will be made later of the $SO(9)$ Fierz identity
\eqn\fierz{\calS^A \calS^B = {1\over 2}\delta^{AB} + {1\over 32}
(\gamma_{IJ})^{AB} \calS\gamma^{IJ} \calS + {1\over 96}
(\gamma_{IJK})^{AB} \calS\gamma^{IJK} \calS,} which is valid for
sixteen-component  Majorana spinors.

\subsec{Summary of vertex operators}

  {\bf Graviton Vertex
Operator}:

The transverse graviton vertex operator is given by
\eqn\grav{\eqs{
U_h = h_{IJ}\big(\dot X^I\dot X^J -2\dot X^I \calR^{JM}k_M +
2\calR^{IL} \calR^{JM}k_Lk_M \big), }} while the longitudinal
vertex operators  are given by
\eqn\longrav{U_{h^-} = - h_I^{\ -} p^+ \big(\dot X^I - \calR^{JM}k_M \big),}
\eqn\longlong{U_{h^{--}} = h^{--} p^+ p^+ .}
The graviton wavefunctions in these expressions satisfy the
conditions \gravcon.
\medskip

 {\bf Gravitino Vertex Operator}:

 The transverse gravitino vertex operator is given by
\eqn\tino{
U_{\psi} = \psi_I\sqrt{p^+} \calS\big(\dot X^I-
2\calR^{IJ}k_J\big)
 +{1\over\sqrt{p^+}}
\pt_I\big\{ \gamma\cdot\dot X\calS\big(\dot
X^I-2\calR^{IJ}k_J\big) +{8\over 9}\gamma^L\calS
\calR^{IJ}\calR^{LM} k_J k_M \big\} \; ,} while the longitudinal
gravitino vertex  is given by
\eqn\longravi{U_{\psi^-} = -\psi^- p^+ \calS  -
\pt^- p^+ \gamma\cdot\dot X\calS .} They  satisfy the physical
state conditions \spincons.

\medskip

{\bf Three-Form Potential Vertex Operator}:

The  vertex operator for the transverse components of the field
strength of the potential, $C^{(3)}$, is given   by
\eqn\cthree{
U_{C^{(3)}} =F_{ILMN}\big(\dot X^I-{2\over
3}\calR^{IJ}k_J\big)\calR^{LMN}  .} The  longitudinal components
of the field strength have vertex
   operators defined by
\eqn\longc{U_{C^{(3)-}} =- F_{LMN}^{\ \ \ \ \ \ -} p^+ R^{LMN}  ,}
and the components of $F$ are constrained by \conlc.


\medskip

 One  direct way of checking the expressions for the vertex
operators is to evaluate their matrix elements between on-shell
states and  compare with the  expressions for the three-particle
couplings in the field theory. As an example, consider the
three-graviton vertex in light-cone gravity which is given by
\goroff\
\eqn\cubic{
L_3 \sim 2h^{JL}\p_K h_{IJ}\p_L h^{IK} - h^{KL}\p_K h_{IJ}\p_L
h_{IJ} .}
 This is easily seen to be equal to the expression obtained by taking the
a matrix element of the   graviton vertex operators,
\eqn\amp{
A_3 = \langle h_1, k_1 | h_2^{IJ}\big(p^Ip^J -2p^I\calR^{JM}k_2^M
+2\calR^{IL}\calR^{JM}  k_2^L k_2^M \big) e^{-ik_2\cdot X} |h_3,
k_3 \rangle } where $|h_3, k_3 \rangle  = h_3^{IJ} e^{-ik_3\cdot
X}|IJ\rangle  $ and use has been made of the fact that
 $\calR^{JM}$ acts as an angular momentum  on the graviton
states so that,
\eqn\ang{
\calR^{JM}|RS\rangle  = \half\big(\delta^{MR}|JS\rangle
-\delta^{JR}|MS\rangle + \delta^{MS}|RJ\rangle
-\delta^{JS}|RM\rangle\big) . }

\subsec{Supersymmetry transformation on vertex operators}

The vertex operators  defined in \grav\ to \cthree\  may be
derived from the requirement that they form a representation of
the linearized supersymmetry transformations of eleven-dimensional
supergravity. We will go some way to demonstrating  this
explicitly although we will not give a complete discussion of all
the supersymmetry transformations.

Proceeding by analogy with the method used in light-cone gauge string
theory \gsc\ (which was explained more fully in \greenseib ),
the 32-component supercharge, $Q^{\hat A}$,  decomposes into the
$SO(9)$ spinors $Q^A$ (with parameter $\eta^A$)  and $\tilde Q^A$
(with parameter $\epsilon^A$).  Whereas the momentum   and  the $Q^A$
supercharge   are given by their free-field expressions the   operator
$\tilde Q^A$, as well as the hamiltonian $H$,
receive  interaction corrections at every order in perturbation theory.
The algebra,
\eqn\susyal{\eqs{\{ Q^A, Q^B \} =2p^+\delta^{AB}, \quad \{\tilde Q^A, Q^B\}
&=\sqrt{2}\gamma_I \dot X^I, \quad \{\tilde Q^A, \tilde Q^B\}=2
H\delta^{AB}, \cr   [H, Q^A] = 0 &= [H, \tilde Q^A] ,}}
determines the form of the interaction corrections.

We may expand the interaction  operators in powers of the fields,
$H = H_2 + H_3 + \dots$, $\tilde Q^A = \tilde Q_2^A +  \tilde
Q_3^A + \dots$, where a subscript ${}_n$ indicates an operator
that has $n$ powers of the fields.    The lowest-order corrections
to the free-field algebra
  are given by
\eqn\susylin{[H_3, Q^A] =0, \qquad \{\tilde Q_3^A, Q^B\} = 0,}
and
\eqn\susynon{[H_2, \tilde Q_3^A] + [H_3, \tilde Q_2^A] =0,  \quad
\{\tilde Q_3^A, \tilde Q_2^B\} +  \{\tilde Q_2^A, \tilde Q_3^B\} =
2H_3\delta^{AB}.  }
These expressions determine how the free-field generators  ($H_2$,
$Q^A$ and $\tilde Q_2^A$)  transform the interaction terms.

To make contact with the vertex operator description we can associate
each $n$-field interaction term  with a  state in the $n$-particle
Hilbert space so that
\eqn\states{H_3 \to | H\rangle_3, \qquad \tilde Q_3^A \to | \tilde
Q^A\rangle_3.}
Cubic interaction vertices are then identified with  matrix elements
in which one of the three external legs is a physical on-shell state,
$\Phi$,
\eqn\vertdef{V_\Phi = \langle \Phi  | H\rangle_3, \qquad  W_\Phi^A =
\langle \Phi | \tilde Q^A\rangle_3 .}
These  are  representations of the interactions in a two-particle
space that can  be represented by single creation and annihilation
operators in a  standard manner.
The supersymmetry transformations of the vertices therefore follow by
taking the  corresponding matrix element of  \susynon,
\eqn\cosu{\eqalign{
\delta V_{h}&= V_{\psi}(\delta \psi), \quad \delta V_{\psi}=
V_{h}(\delta h)+ V_{C^{(3)}}(\delta C^{(3)}), \quad \delta
V_{C^{(3)}}= V_{\psi}(\delta \psi) \cr
\tde V_{h}&= V_{\psi}(\tde \psi)+\epsilon^A\dtau W_h^A, \cr
\tde V_{\psi}&= V_{h}(\tde h)+ V_{C^{(3)}}(\tde
C^{(3)})+\epsilon^A\dtau W_\psi^A,  \cr
\tde V_{C^{(3)}}&= V_{\psi}(\tde \psi)+\epsilon^A \dtau W_{C^{(3)}}^A \; ,
}}
where the $\delta$'s are the variations associated with the free-field
supersymmetry operators in \susylin\ and \susynon\ and the time
derivatives are generated by $H_2$.   The operators $W_h$, $W_\psi$
and $W_{C^{(3)}}$, defined in \vertdef, transform under supersymmetry as
\eqn\wtrans{\eqalign{&\delta W_h = W_\psi(\delta \psi), \quad
   \delta W_\psi = W_h(\delta h) + W_{C^{(3)}} (\delta C^{(3)}) , \quad
\delta W_{C^{(3)}} =  W_\psi (\delta \psi), \cr
&\epsilon^A_1\tde_{\epsilon_2}W^A_h -
\epsilon^A_1W^A_\psi(\tde_{\epsilon_2}\psi ) -( 1 \leftrightarrow
2 ) = 2 \epsilon^A_1\epsilon^A_2 V_h,  \cr
&\epsilon^A_1\tde_{\epsilon_2}W^A_\psi -
\epsilon^A_1\big(W^A_h(\tde_{\epsilon_2}h)+W^A_{\ast}(\tde_{\epsilon_2}\ast)
\big) -( 1 \leftrightarrow 2 ) = 2 \epsilon^A_1\epsilon^A_2
V_\psi,  \cr &\epsilon^A_1\tde_{\epsilon_2}W^A_{\ast} -
\epsilon^A_1W^A_\psi(\tde_{\epsilon_2}\psi ) -( 1 \leftrightarrow
2 ) = 2 \epsilon^A_1\epsilon^A_2 V_\ast  . }}
  In \cosu\ and
\wtrans\ the supersymmetry-transformed wavefunctions such as
$\delta h$ are given by the linearized supersymmetry
transformations of the fields, \lcgrav, \lcrs\ and \delcc.


\subsec{Linear supersymmetry on the graviton vertex  }

Under the linearly realized supersymmetry  transformation, $\delta
X^I = 0$, $\delta\calS = \sqrt{p^+} \eta $,  the graviton vertex
operator, $V_h = U_he^{ik\cdot X}$ (with $U_h$ defined in \grav),
transforms according to,
\eqn\vgrav{\eqs{
\delta V_h&= - h_{IJ}\sqrt{p^+}\big(\dot{X}^I\eta  \gamma^{JK}\calS\;
 k_K +\half \eta \gamma^{IK}\calS\;\calS \gamma^{JL}\calS\; k_K k_L
\big) e^{-ik \cdot X}  \cr & =
k_{[L}h_{I]J}\eta\gamma^{LI}\sqrt{p^+}\calS \big(\dot X^J -2 \calR
^{JM} k_M \big) e^{-ik \cdot X} .}} Hence the linear supersymmetry
transformation of the graviton vertex operator is the gravitino
vertex, $U_\psi e^{-ik\cdot X}$ (with $U_\psi$ defined by \tino),
with the wavefunction $\delta \psi$ given by \lcrs. None of the
other terms in the gravitino vertex \tino\ contribute since they
depend on $\delta \tilde{\psi}$ \lcrs,  which vanishes for the
linear components of supersymmetry when  $k^+=0$.

\subsec{Nonlinear supersymmetry on the graviton vertex}
 To determine  the full structure of gravitino vertex operator
 including
the terms in \tino\ depending on $\tilde{\psi}$, a non-linearly
realized supersymmetry transformation on $U_h$ has to be
performed. Under the non-linear supersymmetry transformations
$\tilde\delta X^I = -2\epsilon\gamma^I\calS/\sqrt{p^+} \;
,\;\tilde\delta\calS = i\dot X^I\gamma^I\epsilon/\sqrt{p^+}$ . The
transformation of the  graviton vertex is given by
\eqn\trangrav{\eqalign{\sqrt{p^+}\tilde{\delta}V_h =&
 h_{IJ} \big(- \dot{X}^I\; (\tilde{\delta}\calS)
\gamma^{JK}\calS\; k_K  +
{1\over 2} (\tilde{\delta}\calS) \gamma^{IK}\calS\;\calS \gamma^{JL}\calS\;
k_K k_L\big)e^{-ik \cdot X} \cr
&+i\epsilon\g^Ik_I\calS\, U_h e^{-ik \cdot
X}   \cr
=& -i h_{IJ}\epsilon(\gamma^{IL}\gamma^N
-2\delta^{NL}\gamma^I+2\delta^{NI}\gamma^L)\calS
k_L\dot X^N\big(\dot X^J -2\calR^{JM}k_M \big)e^{-ik \cdot X} \cr
&+ 2i h_{IJ}\epsilon\gamma^N k_N \calS \big(\dot X^I\dot
X^J - 2\dot X^I\calR^{JM} k_M + 2\calR^{IL}\calR^{JM}k_L k_M
\big)e^{-ik \cdot X} .
}}
 The last term,
which is of order $\calS^5$, can be re-expressed  using \fierz\ as
\eqn\fierzaf{
i h_{IJ}\epsilon\g^N k_N\calS\;\calR^{IL}\calR^{JM} k_L k_M\;
=  -{2\over 9} i h_{IJ}\epsilon \gamma^{IL} \gamma^K \calS\; \calR^{KM}
\calR^{JN} k_L k_M k_N\;  . }

Hence, \trangrav\ can be written as
\eqn\trangravb{\eqalign{\tilde{\delta}V_h & =
2ik^-h_{IJ}\epsilon\gamma^I\,\sqrt{p^+}\calS\big(\dot X^J -2\calR^{JM}
k_M\big)e^{-ik \cdot X} \cr
&\quad + i k_{[L}h_{I]J}
\epsilon\gamma^{LI}\, {1\over \sqrt{p^+}} \big(\g\cdot\dot
X\calS (\dot X^J -2\calR^{JM} k_M) + {8\over 9}\gamma^K \calS
\calR^{KM} \calR^{JN} k_M k_N  \big) e^{-ik \cdot X}\cr
& \quad   + \epsilon^A {d W_h^A \over d\tau}  ,\cr}}
where
\eqn\whdef{W_h^A =  {1\over \sqrt{p^+}}  h_{IJ}(\gamma^I\calS)^A \big(\dot X^J
-2R^{JM}k_M\big) e^{-ik \cdot X}.}
This is precisely of the expected form.  The right-hand side of
\trangravb\ is the
sum of the   gravitino vertex operator \tino\  with  wave-function
$\tde\psi(\pt)$  and the time derivative of $W_h$, which is thus determined.


\subsec{Linear supersymmetry on the gravitino vertex}

We shall now
consider the linear supersymmetry transformation of the gravitino
vertex operator \tino\ which  contains  much information
concerning the graviton and three-form potential vertex operators.
In fact, since  the transformation contains terms of order
$\calS^0, \calS^2$ and $\calS^4$, which are the same orders as in
$U_h$ \grav\ and $U_{\ast}$ \cthree, it will be sufficient to
calculate the linear supersymmetry transformation to confirm the
structure of the other two vertex operators. The linear
supersymmetry transformation, \lin, of the gravitino vertex is
determined by,
\eqn\susyrs{\eqs{
\delta V_{\psi} =& (\psi_I p^+\eta+\tilde\psi_I\gamma_L\eta\dot{
  X}^L)\big(\dot X^I - 2\calR^{IJ} k_J \big)\expo +(\psi_I p^+
\calS+\tilde\psi_I \gamma_L \calS \dot
{X}^L)(-\eta\gamma^{IJ}\calS k_J)\expo  \cr
&+ {8\over 9}\tilde\psi_I \gamma_L\eta \calR^{IJ}\calR^{LM} k_J k_M\expo
+{8\over 9}\tilde\psi_{(I}\gamma_{L)}\calS\eta\gamma^{LM}\calS
\calR^{IJ}k_J k_M \expo .
}}

In order to check that this agrees with the expected transformation it
is necessary to use Fierz transformations to rearrange these terms so
that  $\eta$  contracts into the gravitino wavefunction $\psi$ or
$\tilde{\psi}$.   We will make use of four identities, ignoring the
longitudinal polarizations,  that follow from \fierz.  Firstly,
\eqn\fierzone{
-\psi_I\calS\calS\gamma^{IJ}\eta k_J = \psi_I \eta \calR^{IJ} k_J
-3\psi_I\gamma^L \eta \calR^{ILM}k_M.}
Secondly,
\eqn\fierztwo{\eqalign{
-\pt_I\gamma^P \calS\calS\gamma^{IJ}\eta \dot X^P k_J \expo =&
-(\pt_I\gamma^L \eta-\pt_L\gamma^I \eta)\dot X^I \calR^{LM}k_M
\expo-\pt_L\g^M \eta  k^{-}p^+ \calR^{LM}\expo \cr
&+{3\over 2}(\pt_I\gamma^{LM} \eta k_N\expo -  2\pt_L\g^{IM}\eta k_N)\dot
 X^I \calR^{LMN}\expo \cr
 &-{3\over 2}\pt_L \g^{MN}\eta k^{-}p^+ \calR^{LMN}\expo
+3\pt_L\eta k_M \dot X^N \calR^{LMN} \expo \cr
&+\dtau \Big\{ \big(i\pt_L \gamma^M\eta \calR^{LM} + {3\over 2}i\pt_L
 \g^{MN}\eta \calR^{LMN} \big)\expo\Big\} .
}}
The third Fierz identity that we shall use is
\eqn\fcalS{
\tilde\psi_I\gamma^L\calS \eta \gamma^{LM}\calS \calR^{IJ}k_J k_M =
{5\over 2}\,\pt_I\gamma^L \eta \calR^{IJ}\calR^{LM}k_J k_M -{9\over
4}\,\pt_I\gamma^{LM}\eta k_N \calR^{IJ}\calR^{LMN} k_J  ,}
while the fourth is
 \eqn\fthetb{\eqs{
\tilde\psi_L\gamma^I\calS\calS\gamma^{LM}\eta \calR^{IJ}k_J k_M
=&-{3\over 2}\,\big( \pt_I\gamma^{LM}\eta k_N -2\pt_L\g^{IM}\eta
k_N\big)\calR^{LMN}\calR^{IJ}k_J \cr &+ 3\,\pt_L\eta k_N
\calR^{LMN}\calR^{MJ}k_J . }}
Substituting these identities into \susyrs\ gives
\eqn\delrs{\eqalign{ \delta V_{\psi}&=  \pt_{(I}\gamma_{J)}\eta(\dot
X^I\dot X^J -2\dot X^I \calR^{JM} k_M +2\calR^{IL}\calR^{JM}k_L
k_M ) \expo \cr 
&+{3\over 2} \left(\pt_I\gamma^{LM}\eta
k_N-2\pt_L\gamma^{IM}\eta k_N\right)\dot X^I \calR^{LMN}\expo
+\dtau\big({3\over 2}i\pt_L\gamma^{MN}\eta \calR^{LMN}\expo\big) \cr
&-\big( {5\over 3} \pt_I\gamma^{LM}\eta k_N - {4\over 3}\pt_L
\gamma^{IM}\eta k_N \big)\calR^{LMN} \calR^{IJ} k_J\expo  \cr 
&+\delta V^\prime  ,
}} 
where $\delta V^\prime$ denotes terms
which contain non-transverse polarizations.

The first line of \delrs\  is   the graviton vertex operator \grav\
with a polarization tensor $\delta h_{IJ}(\eta)$ which was defined in
\lcgrav. The  terms in the second and third lines of \delrs\   have the
right form to reproduce the three-form potential vertex
operator \cthree\  as follows.
Firstly using \fierzone\ and \fierztwo,
\eqn\cthreess{\eqalign{
(\pt_I\g_{LM}\eta k_N-&2\pt_L\g_{IM}\eta k_N)\dot X^I \calR^{LMN}\expo\cr
&=4\pt_{[I}\g_{LM}\eta k_{N]}\dot
X^I \calR^{LMN}\expo -i\dtau(\pt_L \gamma_{MN}\eta \calR^{LMN}\expo
)\; 
}}
Furthermore, using \fcalS\ and \fthetb, the terms in the third line of
\delrs\ can be  written as
\eqn\asttr{-\big( {5\over 3} \pt_I\gamma_{LM}\eta k_N - {4\over
3}\pt_L\gamma_{IM}\eta
k_N \big) \calR^{LMN}\calR^{IJ}k_J\expo  = -4  \pt_{[I}\gamma_{LM}\eta
k_{N]}\calR^{LMN}\calR^{IJ}k_J\expo .}
The terms in the second and third lines of \delrs\ therefore combine into
\eqn\astall{
4\times {3\over 2} \pt_{[I}\gamma_{LM}\eta k_{N]}\big( \dot{X}^I - {2\over
3}\calR^{IJ}k_J \big)\calR^{LMN}\expo ,
}
which is the $C^{(3)}$ vertex operator evaluated with the
supersymmetry transformed wavefunction.


\subsec{Linear supersymmetry on the three-form potential vertex}
   The structure of the three-form pontential vertex operator has
already been determined from the linear supersymmetry
transformation of the gravitino vertex but an additional check is
given by the linear supersymmetry transformation of the $C^{(3)}$
vertex.   Given the expression  for $U_{C^{(3)}}$ in \cthree\ this
vertex  can be written as
\eqn\cthreea{\eqs{
V_{C^{(3)}} &=F_{ILMN}\big(\dot X^I-{2\over
3}\calR^{IJ}k_J\big)\calR^{LMN} \expo\cr &= -3 k_L \ast_{IMN}
\big(\dot X^I-{2\over 3}\calR^{IJ}k_J\big)\calR^{LMN}\expo  +
({\rm total\; derivative})\, . }} In the following   we will drop
the total derivative term, although this is an important
contribution to contact interactions.   The linear supersymmetry
transformation of \cthreea\ gives
\eqn\astsusy{
\delta U_{\ast} = -{\sqrtp\over 2}k_L\ast_{IMN}\eta\gamma^{LMN}\calS
\big(\dot X^I-{2\over 3}\calR^{IJ}k_J\big) +{\sqrtp\over 12}
k_Jk_L\ast_{IMN}\eta\gamma^{IJ}\calS\calS\gamma^{LMN}\calS
}
Using the following  Fierz identity,
\eqn\astfierz{
k_Jk_L\ast_{IMN}\eta\gamma^{IJ}\calS\calS\gamma^{LMN}\calS =
k_Jk_L\ast_{IMN}\big(\eta\g^{LMN}\calS\calS\g^{IJ}\calS +{1\over
3}\calS\g^{LMNIP}\eta\calS\g^{PJ}\calS\big)
}
\astsusy\ can be written as
\eqn\astsusya{
\delta U_{\ast} = -2\sqrtp
k_L\ast_{IMN}\big(\quart\eta\gamma^{LMN}\calS\dot X^I -{1\over
3}\eta\gamma^{LMN}\calS{\cal R}^{IJ}k_J +{1\over
18}\eta\gamma^{LMNIP}\calS {\cal R}^{PJ}k_J\big). }
 It is easy to check that this expression is equal to the part of
$ U_{\psi}(\delta\psi)$ which depends on $\ast$.  Therefore,
 $\delta U_{C^{(3)}}$ is consistent with the
   expected supersymmetry transformation \cosu\ (ignoring the total
   derivatives).


\newsec{Dimensional reduction on $S^1$}

Type IIA supergravity arises by dimensional reduction of
eleven dimensional supergravity on a circle \redMIIA\ so that
 compactification of the  vertex operators
on a circle of radius $R_{11}$ should coincide
with the vertex operators of type IIA
supergravity.  For simplicity,  we will  set the
momentum in the eleventh dimension to zero so that $k_{11}=0$.
The case of non-zero $k_{11}$ describes
interactions of D-particles.  In making this reduction we will
decompose  $SO(9)$ spinors into their $SO(8)$ components so that
\eqn\decos{\calS^A = (S^a, \tilde S^{\dot a}),}
where   $a, \dot a$ label the two inequivalent $SO(8)$ spinors ($a,
\dot a = 1, \dots,8$)  and $\{S^a, S^b\} = \delta^{ab}$,$\{S^{\dot a},
S^{\dot b}\} = \delta^{\dot a\dot b}$.     The $16 \times 16$ $SO(9)$
gamma matrices decompose into the standard $8 \times 8$ matrices,
$\gamma^i_{a \dot b}$ and
$\gamma^i_{\dot a b}$, in the following manner,
\eqn\gamdec{\gamma^I = \pmatrix{0 & \gamma^i \cr
                                      \gamma^i & 0},}
for $I = 1, \dots, 8$, and
\eqn\gamnew{\gamma^{11} \equiv \prod_{I=1}^8 \gamma^I = \pmatrix{1 & 0 \cr
                         0 & -1}.}

The following identities hold when $I,J, K = 1, \dots, 8$,
\eqn\iden{ \calR^{ij}  = {1\over 4}S\gamma^{ij} S + {1\over
4} \tilde S\gamma^{ij} \tilde S,\quad \calR^{ijk}  = {1\over 6}
S\gamma^{ijk} \tilde S, }
 where $i,j,k$ are $SO(8)$ vector
indices, while if one of the transverse  vector  indices is in the
eleventh direction we have
\eqn\elevid{\eqalign{
\calR^{ij\,11} &= {1\over 12} S\gamma^{ij} S  - {1\over 12} \tilde
S\gamma^{ij}\tilde S,  \cr
 \calR^{i\,11} & = {1\over 4} \tilde S \gamma^i S - {1\over 4} S
\gamma^i \tilde S = {1 \over 2} \tilde S \gamma^i S .
}}
The eleven-dimensional Fierz transformation, \fierz, gives the following
well-known ten-dimensional relations,
\eqn\fierzten{
S^aS^b = \half \delta^{ab} +{1\over 16}\gamma_{ij}^{ab}S\gamma^{ij}S,
\qquad \tilde S^{\dot a}\tilde S^{\dot b} =\half \delta^{\dot a\dot b}
+{1\over 16}\gamma_{ij}^{\dot a\dot b}\tilde S\gamma^{ij}\tilde S.
}
To begin with we will  review how the IIA supergravity vertices  arise
from the point-particle limit of those of  the IIA superstring.

\subsec{Point particle limit of the IIA superstring vertices }

The vertex operator for any   massless field, $\Phi$,  in  IIA
superstring theory has the form,
\eqn\genvert{V^{(IIA)}_\Phi = U_\Phi^{(IIA)} e^{-ik \cdot X},}
where
\eqn\genpre{U^{(IIA)}_\Phi = \zeta^\Phi_{{\cal A} {\cal B}} {\cal
O}^{\cal A} \tilde {\cal O}^{\cal B} ,}
and  $\zeta^\Phi_{{\cal A} {\cal B}}$ is the on-shell wavefunction for
$\Phi$ which is either a second-rank tensor, a spinor-vector or a
bi-spinor --- ${\cal A}$ and  ${\cal B}$ may either be vector or
spinor labels.  The prefactor is  correspondingly a product of
left-moving and right-moving  bosonic or fermionic operators.    In
the light-cone gauge  these
are classified into $SO(8)$ representations.

The left-moving and right-moving bosonic operators are  either
transverse vectors,
\eqn\bosoop{{\cal B}^i =  \partial X^i - \half S\gamma^{il} S k_l\, ,
\qquad \tilde {\cal B}^i =  \bar \partial X^i - \half \tilde
S\gamma^{il} \tilde S k_l }
where $i=1,\dots,8$, or   singlets,
\eqn\bosing{ {\cal B}^+ = \tilde {\cal B}^+ =  p^+.}
The left-moving fermionic prefactor  has two pieces corresponding to
the two inequivalent $SO(8)$ spinors,
\eqn\fermone{{\cal F}^a = \sqrt{p^+} S^a\, , \qquad {\cal F}^{\dot a}
= {1 \over \sqrt{p^+}} (\gamma\cdot\partial  X S)^{\dot a}-  {1\over
6 \sqrt{p^+}}  :(\gamma^lS)^{\dot a}S\gamma^{lm}S:k_m, }
while the pieces of the right-moving fermion prefactor are
\eqn\fermtwo{\tilde {\cal F}^{\dot a} = \sqrt{p^+}\tilde S^{\dot a}\,
, \qquad \tilde {\cal F}^a = {1 \over \sqrt{p^+}}(\gamma\cdot\bar
\partial  X \tilde S)^a -{1\over 6 \sqrt{p^+}}
:(\gamma^l \tilde S)^a \tilde S\gamma^{lm} \tilde S:k_m  .}
  In these expressions $\partial X \equiv (\partial_\tau + i
\partial_\sigma) X$, $\bar \partial X \equiv (\partial_\tau - i
\partial_\sigma) X$ and  $\{\tilde S^{\dot a}, S^b\} =0$.
The point-particle limit of IIA string theory is simply obtained  by
dropping all the
$\sigma$ dependence of the variables  so that
\eqn\limii{
\partial X^i = \bar \partial X^i = \dot X^i,}
and   only the zero modes of $S$ and $\tilde{S}$ are retained.

The tensor wavefunctions $\zeta_{ij}$, $\zeta_i^{\ -}$ and
$\zeta^{--}$  describe the massless bosonic fields in the \NSNS\
sector while  the bispinors $\zeta_{ab}\, ,
\zeta_{\dot a \dot b}\, , \zeta_{\dot a b}$ and $\zeta_{a \dot b}$
describe the massless  bosons of the \RR\ sector (where the first
index labels the left-movers and the  second index labels the
right-movers).  Using the physical state  conditions the latter can be
written as,
\eqn\rrfields{\eqs{
\zeta_{ab} &= F_{ij} \gamma_{ab}^{ij} + F_{ijkl} \gamma^{ijkl}_{a b},
\quad \zeta_{\dot a \dot b} = F_{ij} \gamma_{\dot a\dot b}^{ij} +
F_{ijkl} \gamma^{ijkl}_{\dot a \dot b} \cr
\zeta_{a \dot b} &= F_i^{\ -} \gamma_{a \dot b}^i + F_{ijk}^{\ \ \ -}
\gamma^{ijk}_{a \dot b}, \quad
 \zeta_{\dot a b} = F_i^{\ -} \gamma_{\dot  a b}^i + F_{ijk}^{\ \  \
-} \gamma^{ijk}_{\dot a b},}}
where $F_i^{\ -} = k_i C^{(1) - } - k^- C^{(1)}_i $, $F_{ij} =
2k_{[i}C^{(1)}_{j]}$ and $F_i^{\ +} =0$ while $F_{ijk}^{\ \ \ -}$ and
$F_{ijkl}$ are defined similarly in terms of
$C^{(3)}$.
The fermions  are described  by the spinor-vectors $\zeta_{i a}$,
$\zeta_{i \dot a}$, $\zeta^{\ -}_ a$ and $\zeta^{\ -}_{\dot a}$.


\subsec{Reduction of  the graviton vertex operator }

The reduction of the eleven-dimensional graviton vertex operator
\grav\  on a circle  gives rise to the  vertex operators of the
graviton ($h$),  dilaton ($\phi$) and one-form gauge potential
($C^{(1)}$) in IIA supergravity theory. For simplicity, we will here
consider only
the transverse parts of the vertex operators.

The covariant relation between the eleven-dimensional M-theory
metric and the ten-dimensional IIA metric in the string frame is
given by the ansatz \wittena\  (setting the \RR\ vector field to
zero),
\eqn\wansatz{G_{\hmu\hnu}dx^{\hmu} dx^{\hnu} = e^{4/3\phi}(dx_{11})^2+
e^{-2/3\phi} g^{\twoa}_{\mu\nu}
dx^\mu dx^\nu  }
from which the corresponding  metric fluctuations are related by
\eqn\wfluc{
h_{\mu\nu} = h^{\twoa}_{\mu\nu} -{2\over 3}\phi\,\eta_{\mu\nu},
\qquad h_{\ele \ele} = {4\over 3}\phi,
 }
 where
$\mu,\nu=0,1,\cdots,9$.   However, we want to compare the
ten-dimensional and eleven-dimensional theories in their
respective light-cone gauges.   In particular, setting $h_{+-}=0$
in \wfluc\ leads to
\eqn\redund{
h^{\twoa}_{+-} = -{2\over 3}\phi = -\half h_{\ele \ele}, }
   while
the usual ten-dimensional light-cone gauge condition in the type
IIA theory sets $h^{\twoa}_{+-}=0$.  This means that in order to
compare with the usual light-cone vertices of the IIA theory  it
is necessary to perform a gauge transformation in order to
transform away the component $h^{\twoa}_{+-}$ after the
identification \wfluc\ is made.  But the condition
$h^{\twoa}_{+-}=0$ can obviously only be compatible with the
eleven-dimensional condition $h_{+-}=0$ by relaxing the
tracelessness condition, $h^I_I = h^i_i + h_{11\, 11} =0$.  As
remarked earlier, in the kinematic regime $k^+=0$ this
tracelessness condition is not a necessary consequence of the
choice of light-cone gauge, and  we will choose the very
convenient alternative constraint
\eqn\elevvan{h_{11\, 11}=0.}

Upon dimensional reduction the ten-dimensional graviton and
dilaton vertex operators  are both contained in
\eqn\tengrav{\eqs{
U^{10}_{h} =& h_{ij}\big(\dot X^i -\half
S\gamma^{il}Sk_l\big)\big(\dot X^j-\half\tilde S\gamma^{jm}\tilde
S k_m\big)+{1\over 8} h_{\ele\ele}S\gamma^{il}S\tilde
S\gamma^{im}\tilde S k_l k_m   \cr &-{1\over 24}\big(k^ph_{pl}
-\half k_l h^p_p\big)\big(S\gamma^{il}SS\gamma^{im}S +\tilde
S\gamma^{il}\tilde S\tilde S\gamma^{im}\tilde S\big) k_m  \cr =&
h_{ij}\big(\dot X^i -\half S\gamma^{il}Sk_l\big)\big(\dot
X^j-\half\tilde S\gamma^{jm}\tilde S k_m\big) \cr &+{1\over
8}h_{\ele\ele}\big(S\gamma^{il}S\tilde S\gamma^{im}\tilde S
-{1\over 6}S\gamma^{il}SS\gamma^{im}S -{1\over 6}\tilde
S\gamma^{il}\tilde S\tilde S\gamma^{im}\tilde S\big) k_l k_m }}
where the light-cone de Donder gauge condition (the first
condition in \conlc) has been used. Substituting \wfluc\ and
\redund\ in  \tengrav\ gives the string-frame expression,
\eqn\tengrava{\eqs{
U^{\twoa}_{h} =& h^{\twoa}_{ij}\big(\dot X^i -\half
S\gamma^{il}Sk_l\big)\big(\dot X^j-\half\tilde S\gamma^{jm}\tilde S
k_m\big) \cr
&+h^{\twoa}_{+-}\big\{\dot X^i\dot X_i -\half\dot
X_i(S\g^{il}S+\tilde S\g^{il}\tilde S)k_l +{1\over 24}(
S\gamma^{il}SS\gamma^m_iS+\tilde S\gamma^{il}\tilde S\tilde
S\gamma^m_i\tilde S)k_lk_m\big\}.
}}
It is easy to see that the terms in the first line of \tengrava\
have the form of
the point-particle limit of the IIA graviton vertex operator. The
dilaton vertex can be identified as the
trace part of the $h^{\twoa}_{ij}$ vertex.  The presence of the
term with polarization tensor $h^{\twoa}_{+-}$ is
simply a reflection of the fact that
the unphysical $h^{\twoa}_{+-}$ polarization \redund\ is generated
by the dimensional reduction \wansatz\ in the light-cone
gauge. Therefore, the condition \elevvan\ $h_{\ele \ele}=0$
is exactly what is needed to remove the redundant terms in \tengrava\
and obtain the IIA graviton and dilaton vertex operators.

The vertex operator of the one-form gauge potential of the IIA theory  is
obtained from the
graviton vertex operator by the identification $h_{11 \mu} =
C^{(1)}_\mu$.  Taking
one of the indices in \grav\ and \longrav\ to be $11$  gives
\eqn\oneform{\eqalign{
U^{10}_{C^{(1)}}  &= h_{i\, 11}\big[ -\dot X^iS\gamma^j \tilde S
k_j + \quart (S\gamma^{il}S+\tilde S\gamma^{il}\tilde S)
S\gamma^j \tilde S \, k_lk_j\big] \cr  &= -k_jh_{i \,
11}\big[S\gamma^j\tilde S\dot X^i-{1\over
24}(S\g^{ji}\gamma^l\tilde S \tilde S\gamma^{lm}\tilde S -\tilde
S\gamma^{ji}\gamma^l S S\gamma^{lm} S\,)k_m \big] }} where the
Fierz identity for $SO(8)$ spinors
\eqn\idenfier{
\zeta^i\tilde S\g^mSS\g^{il}Sk_lk_m
=-{1\over 6}\zeta^i\tilde S\g^{il}\g^pS S\gamma^{pq}Sk_lk_q }
has been used. This coincides with the expression for the one-form
vertex operator of the IIA theory up to a total derivative which
is irrelevant here.


\subsec{Reduction of the gravitino vertex operator }

The eleven-dimensional
gravitino $\Psi_\hmu$ decomposes into  two ten-dimensional
Majorana-Weyl spinor-vectors  $\Psi_{L\, \mu}$, $\Psi_{R\, \mu}$, 
\eqn\spincon{
\Psi_{L\, \mu}  = \half (1 + \Gamma^{11}) \Psi_{\mu} ,\qquad  \Psi_{R\, \mu}
 = \half (1 -  \Gamma^{11})\Psi_\mu
}
where the subscripts $L$, $R$
correspond to the two chiralities  which are correlated in IIA
string theory with the left and right directions on the
worldsheet.

Upon compactification, the above decomposition is accompanied with a
shift of the  eleven-dimensional gravitino which is given by the
covariant relation, 
\eqn\fwitten{
\Psi_{\mu} = \Psi^{\twoa}_{\mu} -\half
\Gamma_{\mu}\Gamma_{11}\Psi_{11} ,}
 where $\Psi^{\twoa}_{\mu}$ is
the gravitino wavefunction in the IIA string frame. Given the
eleven-dimensional light-cone 
gauge condition, $\Psi^+=0$, the type IIA theory has nonvanishing
$\Psi_{\twoa}^+$   proportional to $\Psi_{11}$. This is analogous
to the earlier discussion in which we saw that in general
$h^{\twoa}_{+-}\ne 0$ when $h_{+-}=0$. Just as this was remedied
by making the choice $h_{11\, 11}=0$ as a metric constraint, it is
convenient to make the choice $\Psi_{11}=0$ as an alternative to
the eleven-dimensional gamma-tracelessness condition in \condef. In this way
the supersymmetry transformations of section 3 still work with the
trace part of the graviton vertex mapped to the gamma-trace part
of the gravitino vertex.

 The physical state conditions for the gravitino are then
\eqn\rscon{
k^{\hmu}k_{\hmu} =0, \qquad k^{\hmu}\Psi_{\hmu} =0, \qquad k\cdot
\Gamma\Psi_{\hmu} = k_{\hmu} \Gamma\cdot\Psi, \qquad \Psi_{\ele}=0
}
which can be rewritten as, in the light-cone gauge with $k^+=0$,
\eqn\rsconlc{\eqs{
k^Ik_I =0, \qquad k^I\psi_I(\pt_I) =0, \qquad \psi_{\ele}=0=\pt_{\ele}\cr
k^I\g_I\pt_{\hmu} =
k_{\hmu}\g^I\pt_I, \qquad k^I\g_I\psi_{\hmu} -k_{\hmu}\g^I\psi_I
=\sqrt{2}(k^-\pt_{\hmu}-k_{\hmu}\pt^-) .
}}

The light-cone components of the spinors, \spincon\ , are obtained as
before by  projecting with $P^{\pm}$,
\eqn\psidefs{
\psi^a_{L\, i} =  (P^+ \Psi_{L\, i})^a, \quad
\tilde \psi^{\dot a}_{L\, i}   =   (P^- \Psi_{L\, i})^{\dot a}, \quad
\psi^{\dot a}_{R\, i} = (P^+ \Psi_{R\, i})^{\dot a}, \quad
\tilde \psi^a_{R\, i} = (P^- \Psi_{R\, i})^a.
 }

Without loss of generality, we will consider only the left-handed
gravitino, $\Psi_{L\, \mu}$. The transverse part  of the dimensional
reduction of the gravitino vertex operator, \tino, gives
\eqn\rsred{\eqalign{
U^{10}_{\psi_L} =& \sqrt{p^+}\psi_i S(\dot{X}^i-\half
\tilde{S}\gamma^{ij}\tilde{S}k_j) + {1\over
\sqrt{p^+}}\pt_i(\gamma\cdot\dot X S-{1\over
6}\g_lSS\g^{lm}k_m )(\dot{X}^i-\half
\tilde{S}\gamma^{ij}\tilde{S}k_j ) \cr
&-\half\sqrtp\psi_iSS\g^{ij}k_j -{1\over 9\sqrtp}\pt_i\tilde S\tilde
S\g^{ij}\tilde SS\g^m\tilde Sk_jk_m \cr
&+{1\over 18\sqrtp}\pt_i\g_lS(S\g^{ij}SS\g^{lm}S+\tilde
S\g^{ij}\tilde S\tilde S\g^{lm}\tilde S)k_jk_m
}}
where the subscript $L$ has been omitted and the following Fierz,
\eqn\fierzso{
\tilde \psi_i\gamma_j S S\gamma_{il}Sk_l ={1\over 3}\tilde \psi_j
\gamma_i S S\gamma_{il} S k_l \, ,
}
has been used which can be derived using \fierzten. The terms in the
first line  of \rsred\ are apparently the vertex operator for the
spin-3/2 states  in the point-particle limit of IIA string
theory from which the  gamma-trace part can be separated as the
dilatino vertex operator. It  would, then, be expected that the
remaining terms in \rsred\ should  vanish. Indeed, using the physical
state conditions, \rsconlc, and the  ten-dimensional Fierz, \fierzten,
it can be shown that
these terms vanish up to total derivative terms.


\subsec{Reduction of the three-form potential vertex operator }

The eleven-dimensional three-form  potential gives rise to both
the \NSNS\ two-form, $B_{\mu\nu}$,  and the \RR\ three-form
potential in ten dimensions. The components of  \cthree\  with
one  index in the eleventh direction results in the \NSNS\
two-form vertex operator upon compactification,
\eqn\astred{\eqs{
U^{10}_B =& \half F_{\ele lmn}\big[ -{2\over 3}{\cal R}^{\ele j}{\cal
R}^{lmn}k_j -3\big(\dot X^l -{2\over 3}{\cal R}^{lj}k_j\big){\cal
R}^{mn \ele}\big] \cr
 =& -\quart F_{\ele lmn} \big[\dot X^l\big(S\gamma^{mn}S-\tilde
S\gamma^{mn}\tilde S\big)+{1\over 3}S\gamma^{lj}S\tilde
S\gamma^{mn}\tilde Sk_j -{1\over 6}S\gamma^{mn}S\tilde
S\gamma^{lj}\tilde Sk_j \big] }} where use has been made of
\iden, \elevid\ and the Fierz identity
\eqn\cident{
k_jF_{\ele lmn}S\gamma^j\tilde S\tilde S\gamma^{lmn}S = {3\over
4} k_jF_{\ele lmn}S\gamma^{jl}S\tilde S\gamma^{mn}\tilde S .}
   Up to a total derivative  \astred\ can be written as
\eqn\asttons{
U^{10}_B = \half \ast_{\ele lm}\big[\dot
X^l\big(S\gamma^{mn}S-\tilde S\gamma^{mn}\tilde S\big)k_n+{1\over
2}S\gamma^{lj}S\tilde S\gamma^{mn}\tilde Sk_jk_n \big]. }
 Using  the identification $\ast_{\ele lm} =
B_{lm}$ this expression coincides   with the the point particle
limit of the IIA \NSNS two-form vertex.

The \RR\ three-form potential vertex is obtained by taking all
the indices of \cthree\ to be in the transverse $SO(8)$
\eqn\mthree{
U^{10}_{C^{(3)}} =  {1\over 6} F_{ilmn}\big(\dot {X}^i -{1\over
6}S\gamma^{ij}Sk_j -{1\over 6}\tilde S\gamma^{ij}\tilde Sk_j
\big)S\gamma^{lmn}\tilde S\, . }
     After a few manipulations this is seen to be
 equivalent to the vertex of IIA supergravity,
\eqn\tathree{
U^{\twoa}_{C^{(3)}} = F_{ilmn}\big[ S\gamma^{ilmn}\gamma^p\tilde
S\big(\dot X^p -{1\over 6}\tilde S\gamma^{pq}\tilde{S}k_q\big) +
\tilde{S}\gamma^{ilmn}\gamma^p S\big( \dot X^p-{1\over
6}S\gamma^{pq}Sk_q \big)\,\big] }
   The superscript $^{\twoa}$
indicates that the vertex is that of  IIA supergravity (while the
superscript $^{10}$ in \mthree\ means the vertex is obtained from
dimensional reduction of the appropriate eleven-dimensional
vertex). 


\newsec{One-loop amplitudes in the light-cone gauge}

In this section  the first-quantized description of the
eleven-dimensional theory developed in the previous sections will
be applied   to the  calculation of one-loop Feynman diagrams
 by integrating over the  world-lines of the circulating particles.
This will fill in some of the details  of the calculations of the $R^4$ and
$\lambda^{16}$
 effective interactions of the type IIB theory outlined in \refs{\ggv} and
 \refs{\ggk}.  In addition other interactions of the same dimension
are obtained from one-loop processes in the eleven-dimensional
theory.

The world-line path integral for the one-loop $n$ particle
scattering amplitude in eleven dimensional supergravity is given
by
\eqn\pathint{A_n = \int {dT\over T} \int {\cal D}X \int {\cal
D}S e^{-\int_0^T dt({1\over 2} \dot{X}^2+i S\dot{S})}
\prod_{r=1}^n \Big(\int dt^{(r)} V^{(r)}(t^{(r)})\Big),  }
 where the vertex
operators representing  the emission of  on-shell particles
were defined in section 3. The one-loop amplitude of
eleven-dimensional supergravity compactified to $(11-d)$
dimensions on a $d$-torus,
 $T^d$,  can be written as
\eqn\worldpath{A^{(d)}_n= {1\over \calV_d}\int  {dT\over T}\int
d^{11-d}{\rm\bf p} 
\sum_{\{l_I\}}e^{-T({\rm\bf p}^2+ G^{(d)IJ}l_Il_J)}
\Tr\left\langle \prod_{r=1}^n \big(\int dt^{(r)}V^{(r)}(t^{(r)})\big)
\right\rangle,
}
where $G^{(d)}_{IJ}$ is the metric on $T^d$, $l^I$ are  the Kaluza-Klein
momenta, $\calV_d$ is the volume of the torus $T^d$ and ${\rm\bf p}$
is the loop momentum transverse to the compact 
directions.
The overall factor $1/{\cal V}_d$ is the measure
 for the summation over the Kaluza-Klein momenta
and the trace  in \worldpath\ is taken over the fermionic  modes
$S_A$. 
The brackets denote the path ordered expectation value of the vertex
operators evaluated with the following Green functions involving the
fluctuating quantum fields.
\eqn\greens{\eqalign{
\langle X^i(t)X^j(t^\prime)\rangle=&
\delta^{ij}G_B(t,t^\prime)= \delta^{ij}\big( {|t-t^\prime|\over 2}+
{(t-t^\prime)^2\over 2T}\big) \cr
\langle\calS^A(t)\calS^B(t^\prime)\rangle=&
\delta^{AB}G_F(t,t^\prime)= \delta^{AB}\big( {1\over
2}{\rm sgn}(t-t^\prime)- {t-t^\prime\over T}\big) \; .
}}
The universal momentum factors of $e^{-ikX}$ in the vertex
operators
give contributions to the expectation value of the form
\eqn\expenerg{\langle \prod e^{-ik_iX(t_i)}\rangle = e^{-\sum_{i\neq
j}k_ik_j  G_B(t_i,t_j)}.}
 In the following example  we will only be
interested in the leading terms in the
low energy expansion, which means that only the loop amplitude with the lowest
number of momenta needs to be  considered.  Hence  the contribution
 \expenerg\  can be replaced by $1$ in this approximation.

The $SO(9)$ fermionic spinor operator $\calS$ can be expressed in
terms of eight creation and eight annihilation operators  which
span the space of 256 polarizations of the supergraviton. The
only nonzero contributions come from traces with at least  sixteen
insertions of the operator
 $\calS^A$.   We shall focus on a `protected'
 class of amplitudes where the path integral over the
fermionic world-line variables in which the vertex operators introduce
precisely sixteen fermionic zero modes.    For this
special class of amplitudes there are no contractions involving
the Green functions \greens.  
The bosonic vertex operators \grav--\longc\   contain
the two bilinears,
\eqn\rdefine{\calR^{LMN} = {1\over 12}\calS\gamma^{LMN}\calS,\quad
\calR^{IL} = {1 \over 4}\calS\gamma^{IL}\calS \; . }
 The sixteen factors of $S^A$ in the trace are provided by the  
eight $\calR$'s  in any of the protected  one-loop
amplitudes with external bosons. 
 The trace of  sixteen ${\cal S}^A$  is given by
\eqn\trixts{\Tr\big({\cal S}^{A_1}\cdots
{\cal S}^{A_{16}}\big)=\epsilon^{A_1\cdots A_{16}} \; ,}
 from which it is easy to deduce the tensors that arise from matrix
 elements of the various different
 combinations of $\calR^{LMN} $ and $\calR^{IL}$. 
   For example, the tensor $t_{16}$ defined by the trace involving
 eight $\calR^{IL}$ is defined by
\eqn\tsixtdef{t_{16}^{I_1 I_2 \cdots I_{16}}=\epsilon^{A_1\cdots A_{16}}
\gamma^{I_1I_2}_{A_1A_2}\gamma^{I_3I_4}_{A_3A_4}\cdots
\gamma^{I_{15}I_{16}}_{A_{15}A_{16}} .}
    Nontrivial relations between the various tensors can be found using
Fierz transformations.
Amplitudes with external fermions also contain the trace of sixteen
factors of ${\cal S}^A$ which can be arranged by Fierz tranformations
into the trace of products of eight bilinears of the form \rdefine. 
 
We therefore see that for the protected processes
 the trace over the ${\cal S}^A$'s simply produces a kinematic factor.  The
 leading term in the low energy expansion of the amplitude can then
be  obtained by setting the external momenta to zero in the
 integrand which makes  the integration over the proper times
$t^{(r)}$ trivial, simply giving a power $T^n$ in the integrand. 
The resulting expression for the leading
 term in the low energy expansion of \worldpath\ for a protected
 process with $n$ external states has the form
\eqn\oneloop{
A^{(d)}_n =  {1\over {\cal V}_d} \tilde K_n \int_0^\infty {dT\over
T} T^{n+d/2-11/2} \;
\sum_{\{l_I\} } e^{  G^{(d)IJ} l_I l_J } F({G^{(d)IJ}, l_I})
 }
where $\tilde K_n$ is the kinematic factor for a specific 
scattering amplitude depending on particle species and  $F({G^{(d)IJ},
l_I})$ is a generic function of the torus metric and the Kaluza-Klein
momentum. The extra factor of $T^{d/2-11/2}$ comes from the
integration over the non-compact loop momentum ${\rm\bf p}$. 

An important feature of the protected class of loop amplitudes that we are
considering is that the contractions between the vertex operators does
not generate any factors of the Green functions \greens, 
which have short distance singularities.   
The only contractions are between factors of $e^{ik\cdot X}$.  This
feature is directly related to the fact that the contact interactions
that are present in the complete theory do not, in fact, contribute to
these particular processes.  This justifies the  fact that we have ignored such
interactions.


\subsec{Compactification on a circle}
 We will now review the way in which one-loop  calculations in
 compactified
 eleven-dimensional supergravity  compare with
known results from string theory. The simplest case is that of
compactification   on a circle which is equivalent to IIA string
theory in ten dimensions \wittena.  for simplicity only four-point
functions  involving four bosonic fields in the \NSNS\ sector (the
graviton, dilaton  and two-form potential) will be considered.

 The dimensionally-reduced vertex operators of relevance are given by
\tengrav\ for the graviton and
\asttons\ for the two-form potential.  A
 non-vanishing trace over the fermionic zero modes
only arises from the part of the vertices of order
$S^2\tilde{S}^2$ which are given by
\eqn\relvert{V_h= {1\over 4} h_{ij}k_kk_l
S\gamma^{ik}S\tilde{S}\gamma^{jl}\tilde{S},\qquad V_B=  {1\over 4} B_{ij}k_kk_l
S\gamma^{ik}S\tilde{S}\gamma^{jl}\tilde{S}.}
In \ggv\ the four-graviton one-loop
amplitude in M-theory  compactified on
$S^1$ was calculated and it reproduced the well-known tree and one
loop terms,
\eqn\oned{A^{(1)}_{R^4}
=C   \tilde K+   2 \tilde K
\int_0^\infty d\hat \tau \hat \tau^{ 1/2}
\sum_{\hat l_1 >
0} e^{- \pi  \hat \tau\hat l_1^2 R_{11}^2} =C \tilde K +  \tilde K  \zeta(3)
{1\over R_{11}^3}.}
The value of the constant $C=2\pi^2/3$ can be determined by connecting it,
via T-duality on a further circle, with the IIB theory.
Using the relation $R_{11}= g^{2/3}_s$ (where $g_s$ is the type
IIA  string coupling)   the first term on the right-hand side of
\oned\ is interpreted as  a one-loop term in type IIA string
theory  whereas the second term corresponds to a tree-level string
theory term.

The kinematic factor $\tilde K$  takes the  form,
\eqn\kininvK{\tilde  K = t^{i_1\cdots i_8} t_{j_1\cdots j_8}
\bar{R}^{j_1j_2}_{i_1i_2}\cdots\bar{R}^{j_7j_8}_{i_7i_8} } where
the tensor $t^{i_1\cdots i_8} (i_n=1,\cdots, 8)$ is defined in
\gs\ and the  linearized Riemann tensor  is replaced by a
generalized tensor which also contains the two-form potential and
the  dilaton,
\eqn\ourrhat{\bar{R}_{i_1i_2j_1j_2} =k_{i_1}k_{j_1}
( h_{i_2j_2} +B_{i_2j_2}-{1\over 4} \delta_{i_2j_2}\phi) . }
  This
is the tensor that can be  obtained from the four-point tree-level
amplitude of type II string theories including the \NSNS\ two-form
potential and the dilaton \gsl.


\subsec{ Compactification on a two torus and type IIB string theory }

The compactification of eleven-dimensional supergravity on a two
dimensional torus $T^2$ of volume ${\cal V}_2=R_{8}R_{11}$(in the
M-theory frame) and complex structure $\Omega$ is related to the
IIB string theory compactified on a circle of circumference
$r_B$(in the string frame) \schwarzaspinwall, where the complex
structure is identified as the modulus, $\tau_B$, of IIB string
theory. The $SL(2,Z)$ self-duality of IIB string theory is
identified with the invariance of M-theory under the group of
large diffeomorphisms of $T^2$.

The correspondence between the parameters of the two theories is
\schwarzaspinwall\
\eqn\para{
\Omega = \tau_B (\equiv C^{(0)} +ie^{-\phi_B} ), \quad r_B ={\cal
V}_2^{-3/4}\Omega_2^{-1/4} =R_8^{-1}R_{11}^{-1/2}
}
where $C^{(0)}, \phi_B$ are the IIB \RR\  scalar and dilaton,
respectively, and $\Omega\equiv \Omega_1 +i\Omega_2$.
The metric on the torus is given by
\eqn\tormet{
G^{(2)}_{ij}= {{\cal V}_2\over \Omega_2}\pmatrix{
|\Omega|^2 &\Omega_1\cr \Omega_1&1 },\quad i,j=8,11.
}
The zweibein in a special Lorentz gauge can be written as
\eqn\zweibein{
e_i^a=\sqrt{{\cal V}_2\over \Omega_2}\pmatrix{\Omega_2 & \Omega_1 \cr 0& 1}
}
where  $i, a(=1,2)$ denote the two-dimensional world and  tangent
space indices respectively. The zweibein
parameterizes the coset $SL(2,R)/U(1)$ where the $SL(2,R)$ acts by matrix
multiplication from the
left and the local $U(1)$ acts from the right which can  be used to
fix the gauge \zweibein. Fixing the local $U(1)$ symmetry  leads to
the standard
nonlinear realization of the $SL(2,R)$ transformation acting on  the complex
structure $\Omega$,
\eqn\modular{
\Omega \rightarrow {a\Omega +b\over c\Omega +d} } where the
coefficients are integers and satisfy $ad - bc =1$. It is
convenient to go to a complex basis in the tangent space,
$z=x_{1}+ix_2, \bar{z}= x_{1}-ix_{2}$ in which  the zweibein now
reads
\eqn\zweibeinc{
e_i^a=i\sqrt{{\cal V}_2\over
\Omega_2}\pmatrix{\bar{\Omega}& -\Omega\cr 1&-1},\qquad a=z,\bar{z}}
and the inverse zweibein is
\eqn\zweibeininv{
e^i_a = {1\over 2\sqrt{{\cal V}_2
\Omega_2}}\pmatrix{1 & -\Omega \cr 1& -\bar{\Omega} },\qquad
{a}=z,\bar{z}\; .}

We will begin by reviewing  the one-loop calculations of
four-graviton \ggv\ and sixteen-fermion \ggk\ scattering
amplitudes.
Amplitudes that involve  $\ast$  will also be
calculated and  may be of interest in relation to the matrix
theory calculation of the three-form potential scattering
amplitudes \plefka. Applying \worldpath\ to the case of the  two
torus compactification, the four-graviton one-loop amplitude \ggv\
can be written  as
\eqn\fourgrav{
A^{(2)}_{R^4}   = {1\over {\cal V}_2} \tilde K_{R^4}\sum_{m,n}\int^{\infty}_0
{dT\over T^{3/2}} e^{-T|m+n\Omega|^2/{\cal V}_2\Omega_2} }
where the kinematic factor $\tilde K_{R^4}$ is \kininvK\ with the
two-form pontetial and the dilaton set to zero. A double
Poisson resummation turns  the sum over the Kaluza-Klein charges
$(m, n)$ into a sum over the windings $(\hm, \hn)$ of the loop
around the two cycles of $T^2$ and \fourgrav\ can be expressed as
\eqn\fgrav{
{\cal V}_2 A^{(2)}_{R^4}
 = {2\pi^2\over 3} \tilde K {\cal V}_2 + \tilde K {\cal V}_2^{-1/2}
f^{(0,0)}(\Omega,\bar \Omega)
}
where  $f^{(0,0)}(\Omega,\bar \Omega)$ is the $SL(2,Z)$-invariant
 nonholomorphic modular function,
\eqn\foo{
f^{(0,0)}(\Omega,\bar \Omega)= \sum_{(\hm,\hn)\ne (0,0)}
{\Omega_2^{3/2} \over | \hm + \hn \Omega|^3}. }
The $R^4$ term conserves the  $U(1)$ charge of classical IIB
supergravity and in the
decompactification limit ${\cal V}_2\rightarrow \infty $ 
the first term in \fgrav\ gives a finite  $R^4$ interaction in the
eleven-dimensional theory. 
Other  IIB four-point  bosonic amplitudes 
which conserve the $U(1)$ charge  and
share the structure \fgrav\  
have  kinematic factors that can be schematically written
as\foot{Such terms were also discussed in \kehagias}
\eqn\others{
(\partial \hat F_5)^4 +(\partial G \partial G^*)^2+( \partial P \partial P^*)^2
+ R^2(\partial G \partial G^* + \partial P \partial
P^*)+\partial P
\partial P^*\partial G \partial G^* +\cdots,
}
where the ellipses denote additional mixed terms. Here $\hat F_5$ is
the self-dual five-form field strength and $G$ is 
a complex linear combination of the  \NSNS\ and 
\RR\ three-form field strengths  given by 
\eqn\gdefin{
G = {1\over \sqrt{2\tau_2}} ( {\rm d} C^{(2)} -\tau {\rm d}B  ),
}
 while the Maurer-Cartan form is given by,
\eqn\pdef{
P_\mu  =  {i\over 2}{\partial_\mu\tau\over\tau_2} .}
The fields $\hat F_5$, $G$ and $P$ have $U(1)$ charges  zero, one and two,
  respectively, while the complex conjugate quantities, denoted  $G^*$ and
$P^*$, have   $U(1)$ charges of the opposite sign.

Such terms can be systematically described in linearized IIB
superspace in terms of a constrained scalar superfield,
$\Phi(x,\theta)$, where $\theta$ is a  sixteen-component complex
Grassmann coordinate 
that transforms as a Weyl spinor of $SO(9,1)$.  After imposing the appropriate
constraints $\Phi$ has the expansion \howewest,
\eqn\phidef{
\Phi =\tau_0 + \Delta,}
   where $\tau_0$ is the constant value of the complex scalar
   field and $\Delta$ is the fluctuation defined by
\eqn\deldef{\eqalign{
\Delta  =& (\tau-\tau_0) +i\thbar^*\lambda+
   \bar{\theta}^*\Gamma^{\mu\nu\rho}\theta 
G_{\mu\nu\rho}+i\thbar^*\Gamma^{\mu\nu\rho}\theta\thbar^*\Gamma_{\mu}\p_{\nu}\psi_{\rho}+ \thbar^*\Gamma^{\mu\nu\eta}\theta\thbar^*\Gamma_{\eta}^{\
\rho\sigma}\theta R_{\mu\nu\rho\sigma}  \cr
&+\thbar^*\Gamma^{\mu\nu\rho}\theta\thbar^*\Gamma^{\sigma\eta\omega}\theta
\p_{\mu}(\hat F_5)_{\nu\rho\sigma\eta\omega} +\theta^5 \p^2 \psi^*_{\mu} 
+\bar{\theta}^*\gamma^{\mu\nu\rho}\theta\bar{\theta}^*\gamma^{\sigma}\theta
\bar{\theta}^*\gamma^{\omega}\theta( \p_\sigma \p_\omega
G^*_{\mu\nu\rho}) \cr &+\theta^7\p^3\lambda^* +\theta^8\partial^4\tau^*  
}}
where $\lambda, \psi_{\mu}$  are the dilatino and the
gravitino,
respectively. The terms of order  $\theta^5$  and higher have been written
schematically (except the term of order  $\theta^6$  which will be used
 later).
The linearized interactions of the component fields in the limit
of weak coupling ($\tau_0 \to \infty$) are contained
in
\refs{\ggk}
\eqn\linaction{
S = (\alpha')^3 {\rm Re}\int d^{10}x d^{16}\theta F[\Phi] .}
  The
component interactions are obtained by Taylor expanding $F$ in
powers of $\Delta$  followed by integration over the 16 Grassmann
parameters.
 Of course, this expression does not capture the
  modular properties  of the complete nonlinear interactions.

 The $R^4$ term occurs at  order $\Delta^4$ together with the terms in
 \others.  For example, $(\partial G \partial G^*)^2$ is
 obtained by combining  two powers
 of each of the
$\theta^2$ and $\theta^6$ terms. The Grassmann integration
generates the pattern of contractions between the tensor fields.  The terms in
\others\ also arise naturally from the eleven-dimensional
perspective with appropriate identifications of the polarization
tensors as will be seen shortly.

The sixteen-dilatino term $\lambda^{16}$ has $U(1)$ charge $24$.
This interaction was obtained in \ggk\ by a  covariant argument
using the fact that the zero-momentum eleven-dimensional
gravitino  vertex can be expressed in terms of the
eleven-dimensional supercharge. The same result can be  obtained
using  the light-cone gauge gravitino vertex operator \tino. The
only contributions come from factors in which there are  eight
vertices from each of the $\psi_I$ and $\pt_I$ terms in \tino.
This follows from the fact that  only terms with no net power of
$p^+$ contribute to the momentum integral. Furthermore, there
must be a total of sixteen powers of ${\cal S}$ so that   the
only  terms in the vertex \tino\ that contribute to the amplitude
are
\eqn\rela{
U_{\psi} = \sqrtp\psi_I{\cal S}\dot X^I, \qquad U_{\pt} ={1\over
\sqrtp}\pt_I\gamma\cdot\dot X{\cal S}\dot X^I. }
 The complex dilatino  field $\lambda$ is a ten-dimensional Weyl spinor that
is  identified with a particular polarization state of the
eleven-dimensional gravitino compactified on $T^2$ so its vertex
operator follows simply from \rela\ \refs{\ggk}. The amplitude
can then be obtained in much the same way as for the
four-graviton case  using \worldpath. The significant new feature
is the presence of a complex function of the  Kaluza-Klein
momentum in front of the exponential,
\eqn\sixteen{\eqalign{
{\cal V}_2 A_{\lambda^{16}} &= \tilde{K}_{\lambda^{16}} \sum_{m,n} \int
 {dT\over T} T^{23/2}
 \big({1\over \sqrt{{\cal V}_2\Omega_2}} (m+n\bar \Omega)\big)^{24}
\; e^{-T|m+n\Omega|^2/{\cal V}_2\Omega_2}\cr &=
\tilde{K}_{\lambda^{16}} {\cal V}_2^{-1/2} f^{(12,-12)}(\Omega,
\bar\Omega). }}
  The kinematic factor is $\tilde{K}_{\lambda^{16}}$ is proportional to
  $\epsilon^{A_1A_2\dots A_{16}}\lambda_{A_1} \lambda_{A_2} \dots
  \lambda_{A_{16}}$ 
and $f^{(12,-12)}$ is a  modular form of weight $(12,-12)$ that is
given by
\eqn\ftwofour{
f^{(12,-12)}(\Omega, \bar\Omega) = {\Gamma(27/2)\over
\Gamma(3/2)} \sum_{(\hm,\hn)\ne (0,0)}\Omega_2^{3/2} {(\hm +
\hn\bar\Omega)^{24} \over |\hm + \hn \Omega|^{27}}. }
   In contrast to  the $U(1)$
charge conserving amplitudes considered before, the $\lambda^{16}$
term violates 24 units of $U(1)$ charge and is not present in the
  eleven-dimensional theory. This is seen
explicitly from \sixteen\ which  vanishes in the
eleven-dimensional limit ${\cal V} \to \infty$,   while it
survives in the ten-dimensional IIB limit. This is true for all
the other $U(1)$ charge violating terms. In \sav\ the explicit
form of \ftwofour\ was derived from space-time supersymmetry and
$SL(2,Z)$ invariance of IIB string theory alone.

In \ggk\ the scalar field strength  $P_{\mu}$ \pdef\ of IIB
supergravity  was identified with the component $\omega_{z\mu z}$
of the spin connection of eleven dimensional supergravity on
$T^2$ which can be written as
\eqn\spinc{
\omega_{z\mu z} = \half e^i_z e^j_z \partial_{\mu} h_{ij}, }
where $i,j=8, 11$ and $P^*_{\mu}$ corresponds to
$\omega_{\bar{z}\mu\bar{z}}$. The one-loop amplitude that
generates the interaction $(\p P\p P^*)^2$ in \others\ gets
contributions from the factors $\calR^{zl}\calR^{zm}$ and
$\calR^{\bar{z}l}\calR^{\bar{z}m}$ in the graviton vertex
operator, \grav.

We will now consider the vertex operators  for  the
three-form potential  $\ast$ which reduce, upon  compactification on a torus,
to the IIB two-form potentials and four-form potential $C^{(4)}$. The
correspondence is given by 
\eqn\redast{
\ast_{\mu\nu\rho} =  C^{(4)}_{\mu\nu\rho 8}, \qquad
\ast_{\mu\nu\ele}=B_{\mu\nu}, \qquad \ast_{\mu\nu8} = C^{(2)}_{\mu\nu}
}
 where $\mu,\nu, \rho$  are $SO(8,1)$ indices. 
Therefore, in terms of the M-theory parameters, the complex three-form
field strength $G$ \gdefin\ can be written as
\eqn\ginmc{
k_{[\rho}C^{(3)}_{\mu\nu] z} = {1\over 2\sqrt{{\cal
V}_2\Omega_2}} ( k_{[\rho} C^{(3)}_{\mu\nu]8} -\Omega
k_{[\rho}C^{(3)}_{\mu\nu]11} ) }
  and $G^*$ is identified with
$k_{[\rho}C^{(3)}_{\mu\nu] \bar z}$. The transverse part of the $\ast$
vertex operator 
\cthree\ gives   the vertex for transverse components of the IIB
self-dual five-form field strength  
\eqn\five{
V_{\hat F_5} = k_{[i}\ast_{lmn]}\big( p^i +{2\over
3}\calR^{ij}k_j\big) \calR^{lmn} \expo 
}
and those for the polarization \ginmc\ and its complex conjugate
are given by
\eqn\gvertex{
V_{G} = k_{[l}\ast_{mn]z}\, \big(p_{\bar z}+{2\over
3}\calR^{zj}k_j\big) \calR^{lmn} \expo, \quad V_{G^*}
=k_{[l}\ast_{mn]\bar z}\, \big(p_z+{2\over 3}\calR^{\bar{z}j}
k_j\big) \calR^{lmn} \expo } 
where $i,j,l,m,n$ are $SO(7)$ vector
indices. 

We  will now discuss `protected' interaction terms that
include   the 
 antisymmetric tensor $C^3$ field  associated with the vertices \ginmc\
and \five.
There are two amplitudes that only involve 
 $G$ and ${G}^*$ with nozero $U(1)$ weight namely 
 $G^8$ and $G^5\p^2G^*$. It is easy to see
that in the calculation of the $G^8$ amplitude saturating the
 fermionic modes  picks out the $p_{\bar z}{\cal R}^{lmn}$ term of each
 the vertex $V_G$ in
\gvertex.
For the $G^5\p^2G^*$ amplitude  a combination of four
factors of the $p_{\bar z}{\cal R}^{lmn}$ term in $V_G$, one
factor of the ${\cal R}^{zj}{\cal R}^{lmn}k_j$ term in $V_G$, and
one from the ${\cal R}^{\bar{z}j}{\cal R}^{lmn}k_j$ term in
$V_{G^*}$ is required. These amplitudes can be calculated in the same way as
the $\lambda^{16}$ term and result in effective interactions of
the form $f^{(4,-4)}G^8$ and $f^{(2,-2)} G^5\p^2G^* $, where
\eqn\fothers{\eqalign{
f^{(4,-4)}(\Omega, \bar\Omega) &=  {\Gamma(11/2)\over
\Gamma(3/2)} \sum_{(\hm,\hn)\ne (0,0)}\Omega_2^{3/2} {(\hm +
\hn\bar\Omega)^{8} \over |\hm + \hn \Omega|^{11}}, \cr
f^{(2,-2)}(\Omega, \bar\Omega) &=  {\Gamma(7/2)\over \Gamma(3/2)}
\sum_{(\hm,\hn)\ne (0,0)}\Omega_2^{3/2} {(\hm + \hn\bar\Omega)^4
\over |\hm + \hn \Omega|^7 } .}}
 The transformation properties of the fields
and the generalized modular functions in \fothers\ lead to the
expected  $SL(2,Z)$-invariant interactions in the IIB theory.

More generally, the exact expression for all the protected IIB
higher-derivative interactions that follow from the expansion of
\linaction\  can be deduced by considering the appropriate
decompactification limit of one-loop amplitudes in compactified
eleven-dimensional supergravity.  The resulting interactions are
proportional to
\eqn\generals{
\int d^{10}x \, \sqrt{g} \, e^{-\phi^B/2}\, f^{(w,-w)}(\tau,
\bar\tau){\cal P}^{(2w)} ,}
  where ${\cal P}^{(2w)}$ denotes an interaction between a set of fields
 with net $U(1)$ charge $2w$ and
\eqn\general{\eqalign{
f^{(w,-w)}(\tau, \bar\tau) &= {\Gamma(w+3/2)\over \Gamma(3/2)}
\sum_{(\hm,\hn)\ne (0,0)}\Omega_2^{3/2} {(\hm + \hn\bar\Omega)^{2w} \over
|\hm + \hn \Omega|^{2w+3}} \quad {\rm for} \; w\geq 0 \cr
f^{(w,-w)}(\tau, \bar\tau) &= {\Gamma(|w|+3/2)\over \Gamma(3/2)}
\sum_{(\hm,\hn)\ne (0,0)}\Omega_2^{3/2} {(\hm + \hn\Omega)^{2|w|} \over
|\hm + \hn \Omega|^{2|w|+3}} \quad {\rm for} \; w\leq 0  .
}}

Another  interesting IIB amplitude which can be obtained in this way
  is    the 
amplitude involving four self-dual five-form fields $\hat F_5$. 
 The eleven dimensional vertex  \five\ gives the
amplitude which involves $(\hat F_5)_{ilmn8}$. From  the self-duality of
$\hat F_5$ it follows  that $(\hat F_5)_{ilmn8}$ is equivalent to $(\hat
F_5)_{ijlmn}$, hence the amplitude  calculated in eleven
dimensional supergravity on $T^2$ can be expressed covariantly.
  Indeed, the wave function $\ast_{lmn}$ in \five\ has 35 physical
degrees of freedom 
which is the  number of physical components of  the self-dual
five-form 
$\hat F_5$.  The calculation of the four $\partial \hat F_5$ amplitude 
is analogous to the $R^4$ amplitude and it is straightforward to show
that the amplitude has  the same  
modular structure \fgrav. The main  difference lies in the fact that   the 
$\calR^{ij}\calR^{lmn}$ term in each vertex \five\ contributes  to the
amplitude, resulting in the following kinematic factor 
\eqn\ffivekinf{(\partial \hat F_5)^4=\int d^{16}\theta \Big(\thbar^*\Gamma^{\mu\nu\rho}\theta\thbar^*\Gamma^{\sigma\eta\omega}\theta
\p_{\mu}(\hat F_5)_{\nu\rho\sigma\eta\omega}\Big)^4}
Similarly terms like  $(\partial \hat F_5)R^2$ and $(\partial G \partial G^*)^2$
in \others\ can also be obtained as well
 as terms involving the fermionic fields.
 The $(\partial \hat F_5)^4$ and $(\partial \hat F_5)R^2$ vanish in the
 $AdS_5 \times S_5$ background of IIB in which $\hat F_5$ is a
 constant. This is in accord with expectations based on the AdS/CFT
 correspondence \refs{\juan, \gkpw}.

The $(\partial G \partial G^*)^2$   together with  the $(\partial
\hat F_5)^4$ interaction vertex in nine dimensions, has a piece which
survives the  
decompactification  
limit to eleven dimensions and produces a  $(\partial F_4)^4$ term in
eleven dimensions. This term is a higher derivative correction of the
eleven dimensional supergravity and appears at the same order as
$R^4$.  The four-point scattering 
amplitude of the three-form potential was calculated in \plefka\ using the
matrix theory at one-loop and identified with the classical
supergravity result. It would be of interest to see how the
vertex $(\partial F_4)^4$ together with other terms  (such as the $R^4$
term) arise in matrix theory.

It would be interesting to use the eleven-dimensional vertices in situations with nonzero momentum in the eleventh dimension to describe the interactions of D-particles which should agree with results of \kallosha\  obtained by
quantization of the massive ten-dimensional superparticle in
static gauge, although we have not checked this. Another
interesting generalization   would be the formulation of
covariant superspace vertex operators for the eleven-dimensional
superparticle.  This would be a generalization of the open-string
vertex operators of \gsc.

\medskip
\noindent{\bf Acknowledgements}
\medskip
The work of M. Gutperle  was supported in part by NSF grant PHY-9802484. The
work of H. Kwon was supported in part by ORS Award and Dong-yung Scholarship.
\medskip

\listrefs

\bye